\documentclass{emulateapj}

\def\amin{\ifmmode^{\prime}\else$^{\prime}$\fi}
\def\asec{\ifmmode^{\prime\prime}\else$^{\prime\prime}$\fi}

\shorttitle{The NGC~404 Nucleus}
\shortauthors{Seth et al.}

\begin{document}

\slugcomment{Accepted to ApJ, Mar. 1, 2010}

\title{The NGC~404 Nucleus: Star Cluster and Possible Intermediate Mass Black Hole}

\author{Anil C. Seth\altaffilmark{1,2}, Michele
Cappellari\altaffilmark{3}, Nadine Neumayer\altaffilmark{4}, Nelson
Caldwell\altaffilmark{1}, Nate Bastian \altaffilmark{5}, Knut
Olsen\altaffilmark{6}, Robert D. Blum\altaffilmark{6}, Victor
P. Debattista\altaffilmark{7}, Richard McDermid\altaffilmark{8},
Thomas Puzia\altaffilmark{9}, Andrew Stephens\altaffilmark{8}}

\altaffiltext{1}{Harvard-Smithsonian Center for Astrophysics, 60 Garden Street Cambridge, MA 02138}
\altaffiltext{2}{OIR Fellow, {\tt aseth@cfa.harvard.edu}} 
\altaffiltext{3}{Sub-department of Astrophysics, University of Oxford, Denys Wilkinson Building, Keble Road, Oxford OX1 3RH} 
\altaffiltext{4}{European Southern Observatory, Garching} 
\altaffiltext{5}{Institute of Astronomy, Cambridge University}
\altaffiltext{6}{National Optical Astronomy Observatory, Tucson}
\altaffiltext{7}{Jeremiah Horrocks Institute, University of Central Lancashire}
\altaffiltext{8}{Gemini Observatory, Hilo}
\altaffiltext{9}{Herzberg Institute of Astrophysics, Victoria}

\begin{abstract}

We examine the nuclear morphology, kinematics, and stellar populations
in nearby S0 galaxy NGC~404 using a combination of adaptive optics
assisted near-IR integral-field spectroscopy, optical spectroscopy,
and HST imaging.  
%Resub added
These observations enable study of the NGC~404 nucleus at a level of
detail possible only in the nearest galaxies.
The surface brightness profile suggests the presence of three
components, a bulge, a nuclear star cluster, and a central light
excess within the cluster at radii $<$3~pc.  These components have
distinct kinematics with modest rotation seen in the nuclear star
cluster and counter-rotation seen in the central excess.  Molecular
hydrogen emission traces a disk with rotation nearly orthogonal to
that of the stars.  The stellar populations of the three components
are also distinct, with half of the mass of the nuclear star cluster
having ages of $\sim$1~Gyr (perhaps resulting from a galaxy merger),
while the bulge is dominated by much older stars.  Dynamical modeling
of the stellar kinematics gives a total nuclear star cluster mass of
$1.1 \times 10^7$~M$_\odot$.  Dynamical detection of a possible
intermediate mass black hole is hindered by uncertainties in the
central stellar mass profile.  Assuming a constant mass-to-light
ratio, the stellar dynamical modeling suggests a black hole
mass of $< 1 \times 10^5$~M$_\odot$, while the molecular hydrogen gas
kinematics are best fit by a black hole with mass of $4.5
^{+3.5}_{-2.0}\times 10^5$~M$_\odot$.  Unresolved and possibly
variable dust emission in the near-infrared and AGN-like molecular
hydrogen emission line ratios do suggest the presence of an accreting
black hole in this nearby LINER galaxy.

\end{abstract}

\keywords{galaxies:nuclei -- galaxies: elliptical and lenticular, cD
  -- galaxies: kinematics and dynamics -- galaxies: formation --
  galaxies: structure -- galaxies: individual (NGC~404)}

\section{Introduction}

The centers of galaxies contain both massive black holes (MBHs) and
nuclear star clusters (NSCs).  The presence of MBHs has been
dynamically measured in about 50 galaxies and it appears that most
massive galaxies probably have a MBH
\citep[e.g.][]{richstone98,graham08b}.  Nuclear star clusters are
compact ($r_{eff} \sim 5$~pc), massive ($\sim10^7$~M$_\odot$) star
clusters found at the center of a majority of spirals and lower mass
ellipticals \citep{boker02,carollo02,cote06}.  Unlike normal star
clusters, they have multiple stellar populations with a wide range of
ages \citep{long02,walcher06,rossa06,seth06,siegel07}.  Nuclear star
clusters coexist with MBHs in some galaxies \citep{filippenko03,
seth08, shields08, graham09}.  However, the nearby
galaxies M33 and NGC~205 have NSCs but no apparent central black hole
\citep{gebhardt01,valluri05}, while some high mass core elliptical
galaxies have MBHs but lack NSCs \citep{cote06}.

Occupation fractions and masses for MBHs in lower mass galaxies retain
the imprint of the seed black holes (BHs) from which they form,
information which has been lost due to subsequent accretion in higher
mass galaxies \citep[e.g.][]{volonteri08}.  However, the presence and
mass of MBHs in lower mass galaxies is very poorly constrained.  Most
galaxies are too far away to measure the dynamical effect of a BH with
mass $\lesssim$10$^6$~M$_\odot$ (often referred to as
intermediate mass black holes; IMBHs) with current instrumentation.
Thus the presence of IMBHs in galaxy centers has only been inferred
when AGN activity is observed
\citep[e.g.][]{filippenko89,greene04,satyapal07}.  These AGN provide
only a lower limit on the number of MBHs in lower mass galaxies, and
the BH mass estimates from the AGN are quite uncertain.

Dynamical measurements of MBHs in nearby massive galaxies have
revealed that the mass of a galaxy's central MBH is correlated with
its bulge mass \citep{kormendy95,magorrian98,haring04}.  The scaling
of BH mass with the large-scale properties of galaxies extends to many
measurable quantities including the bulge velocity dispersion
\citep{ferrarese00,gebhardt00,graham08b,gultekin09}.
More recently, \citet{ferrarese06}, \citet{wehner06}, and
\citet{rossa06} have presented evidence that NSCs scale with bulge
mass and dispersion in elliptical and early-type spiral galaxies in
almost exactly the same way as MBHs.  

The similarity between the NSC and MBH scaling relationships led
\citet{ferrarese06} and \citet{wehner06} to suggest that MBHs and NSCs
are two different types of central massive objects (CMO) both of which
contain a small fraction ($\sim$0.2\%) of the total galaxy mass.  This
correlation of the CMO mass with the large scale properties of
galaxies suggests a link between the formation of the two.  However,
the nature of this connection is unknown, as is the relation between
NSCs and MBHs.  Studies of NSCs can help address these issues.  Their
morphology, kinematics, and stellar populations contain important
clues about their formation and the accretion of material into the
center of galaxies \citep[e.g.][]{hopkins10a}.

The scaling relationships of large-scale galaxy properties with the
mass of the CMO might indicate that galaxy centers should be simple
systems.  The Milky Way center clearly shows this not to be the case.
It is an incredibly complicated environment with a black hole
\citep[$M_{BH} = 4 \times 10^6$~M$_\odot$;][]{ghez08} and a nuclear
star cluster \citep[$M_{NSC} \sim 3 \times
10^7$~M$_\odot$][]{genzel96,schodel07,trippe08} that contains stars of
many ages and in different substructures including disks of young
stars in the immediate vicinity of the black hole
\citep[e.g.][]{lu09,bartko09}.  Only through understanding these
complex structures in the Milky Way and finding other examples in
nearby galaxies can we hope to fully understand MBH and NSC formation,
and the links between these objects and their host galaxies.

In our current survey, we are looking at a sample of the nearest
galaxies ($D < 5$~Mpc) that host NSCs.  We are resolving their
properties using a wide range of observational data to understand how
NSCs form and their relation to MBHs.  In our first paper on nearby
edge-on spiral NGC~4244 \citep{seth08b}, we showed evidence that the
NSC kinematics are dominated by rotation, suggesting that it was
formed by episodic accretion of material from the galaxy disk.

This paper focuses on the NSC and possible MBH in NGC~404, the nearest
S0 galaxy.  Table~\ref{proptab} summarizes its properties.  The
stellar populations of the galaxy are predominantly old and can be
traced out to 600\asec~(9~kpc) \citep{tikhonov03,williams10}.
However, HI observations show a prominent nearly face-on HI disk at
radii between 100-400\asec, with detectable HI out to
800\asec~\citep{delrio04}.  CO observations and optical color maps
show the presence of molecular gas and dust within the central
$\sim$20\asec~(300~pc) of the galaxy \citep{wiklind90,tikhonov03},
lying primarily to the NE of the nucleus.  A nuclear star cluster in
the central arcsecond of NGC~404 was noted by \citet{ravindranath01}
from an analysis of NICMOS data.

\begin{deluxetable}{lr}
\tablewidth{\columnwidth} 
\tablecaption{NGC~404 Properties \label{proptab}}
\startdata
\tableline
Distance$^a$ & 3.06 Mpc\\
\hspace{0.2in}$m-M$ & 27.43\\
\hspace{0.2in}pc/\asec & 14.8\\
Galaxy $M_{V,0}$, $M_{I,0}$$^b$ & -17.35, -18.36\\
Bulge/Total Luminosity$^c$ & 0.76\\
Bulge $M_{V,0}$, $M_{I,0}$$^d$ & -17.05, -18.06\\
Bulge Mass$^e$ & 9.2$\times$10$^8$~M$_\odot$\\
HI Gas Mass$^f$ & 1.5$\times$10$^8$~M$_\odot$\\
Molecular Gas Mass$^g$ & 6$\times$10$^6$~M$_\odot$\\
Central Velocity$^h$ & -58.9 km$\,$s$^{-1}$
%\tableline

\enddata

\tablecomments{$(a)$ TRGB measurement from \citet{karachentsev04},
$(b)$ at $r < 200$\asec~from \citet{tikhonov03} corrected for
foreground reddening, $(c)$ for $r < 200$\asec~from \citet{baggett98},
$(d)$ combining above values; we derive a slightly brighter bulge
$M_{I,0} = -18.20$ in \S3, $(e)$ using $M/L_I =
1.28$ from \S5.4, $(f)$ \citet{delrio04}, $(g)$ \citet{wiklind90},
corrected to $D=3.06$~Mpc, $(h)$ heliocentric velocity, see \S4.1.}

\end{deluxetable}

The presence of an AGN in NGC~404 is a controversial topic.  The
optical spectrum of the nucleus has line ratios with a LINER
classification \citep{ho97a}.  Assuming a distance of $\sim$3~Mpc,
NGC~404 is the nearest LINER galaxy; other nearby examples include M81
and NGC~4736 (M94).  Recent studies have shown that most LINERS do in
fact appear to be AGN, with a majority of them having detected X-ray
cores \citep{dudik05,gonzalezmartin06,zhang09}, radio cores
\citep{nagar05}, and many of them possesing mid-IR coronal lines
\citep{satyapal04} and UV variable cores \citep{maoz05}.  However,
NGC~404 is quite unusual in its properties.  No radio core is observed
down to a limiting flux of 1.3~mJy at 15~GHz with the VLA in A array
\citep{nagar05}, however, \citet{delrio04} do detect an unresolved
3~mJy continuum source at 1.4~GHz using the C array.  A compact X-ray
source is detected \citep{lira00,eracleous02}, but its low luminosity
and soft thermal spectrum indicates that it could be the result of a
starburst.  Signatures of O stars are seen in the UV spectrum of the
nucleus, however dilution of these lines suggests that $\sim$60\% of
the UV flux could result from a non-thermal source \citep{maoz98}.
This suggestion is supported by more recent UV observations which show
that the UV emission is variable, declining by a factor of $\sim$3
between 1993 and 2002 \citep{maoz05}.  HST observations of H$\alpha$
show that the emission occurs primarily in a compact source
0.16\asec~north of the nucleus and in wispy structures suggestive of
supernova remnants \citep{pogge00}.  The [\ion{O}{3}] emission has a
double lobed structure along the galaxy major-axis \citep{plana98},
and has a higher velocity dispersion than H$\alpha$ near the galaxy
center \citep{bouchard10}.  The mid-IR spectrum of NGC~404 shows
evidence for high excitation consistent with other AGN
\citep{satyapal04}.  In particular the ratio of the [\ion{O}{4}] flux
relative to other emission lines ([\ion{Ne}{2}], [\ion{Si}{2}]) is
higher than any other LINERs in the \citet{satyapal04} sample and is
similar to other known AGN.  However, [NeV] lines, which are a more
reliable indicator of AGN activity \citep{abel08}, are not detected.
In summary, the case for an accreting MBH in NGC~404 remains
ambiguous, with the variable UV emission providing the strongest
evidence in favor of its existence.

In this paper we take a detailed look at the central arcseconds of
NGC~404 and find evidence that it contains both a massive nuclear star
cluster and a black hole.  In \S2 we describe the data used in this
paper including adaptive optics Gemini/NIFS observations.  We then use
this data to determine the morphology in \S3, the stellar and gas
kinematics in \S4, and the stellar populations in \S5.  In \S6 we
present dynamical modeling from which we derive a NSC mass of $1.1
\times 10^7$~M$_\odot$ and find mixed results on detecting a possible
MBH with mass $<10^6$~M$_\odot$.  In \S7 we discuss these results,
concluding in \S8.

\newpage

\section{Observations \& Reduction}

This paper presents a combination of a wide variety of data including:
(1) high spatial resolution near-infrared IFU spectroscopy with Gemini
NIFS, (2) optical long-slit spectra from the MMT, and (3) HST imaging
data from the UV through the NIR.  Using these data we have determined
the morphology, gas and stellar kinematics, and stellar populations of
the central regions of NGC404.  In this section we describe these
data and their reduction.

\begin{figure*}
\epsscale{1.2}
\plotone{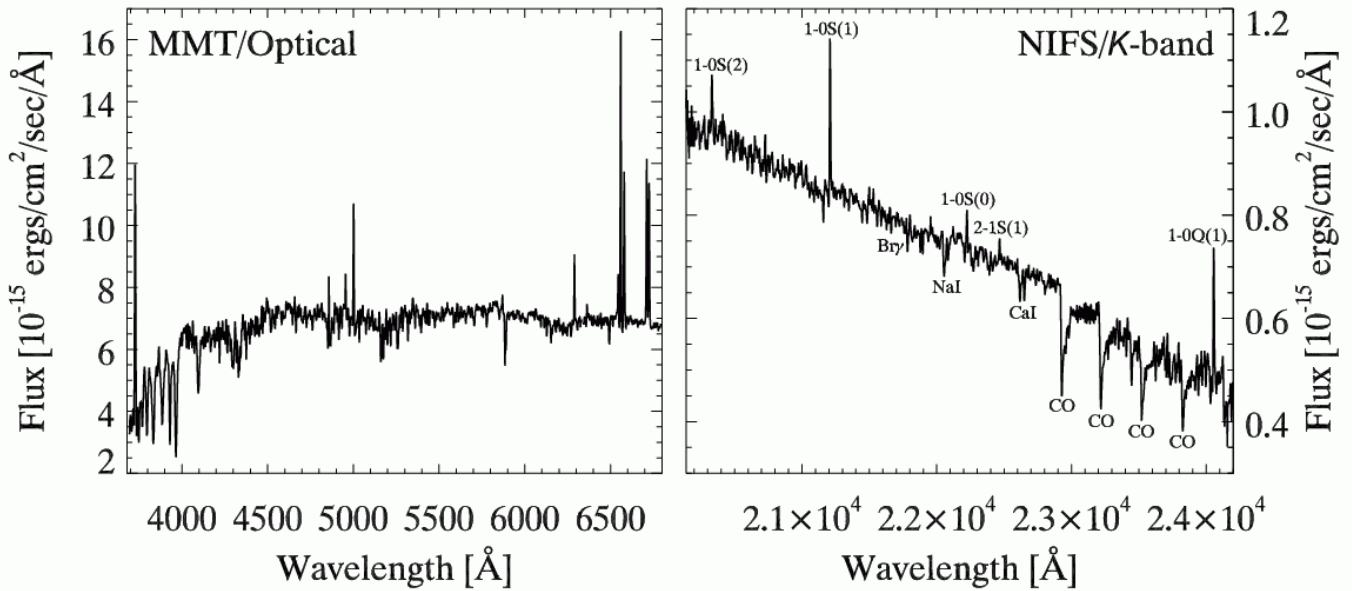}
\caption{{\it Left --} Optical spectrum of the NGC~404 nuclear star
cluster (1\asec$\times$2.3\asec aperture) taken with the blue channel
spectrograph on the MMT.  {\it Right --} $K$-band spectrum of the
NGC~404 nuclear star cluster using a 1\asec~radius aperture from our
Gemini/NIFS data.  Emission lines of H$_2$ are labeled above the spectrum
(see \S3.2), while strong absorption features are labeled below the spectrum (including Br$\gamma$ which is seen weakly in absorption and emission, see Fig.~\ref{halphabrgfig}).
}
\label{optirfig}
\end{figure*}

\subsection{Gemini NIFS data}

Observations of NGC~404 with Gemini's Near-Infrared Integral Field
Spectrometer (NIFS) are part of an ongoing survey of nearby nuclear
star clusters. The data were taken using the Altair laser-guide star
(LGS) system on the Gemini North telescope on the nights of
21st and 22nd September, 2008.  The nucleus itself was used as a
tip-tilt star.  A total of twelve 760 second exposures were obtained;
half were dithered on-source exposures while the others were sky
exposures offset by 131'' to the northwest taken after each on-source
exposure.  A telluric calibrator, HIP~1123, an A1 star, was observed
both nights at an airmass similar to the observations.

The data were reduced using the Gemini v. 1.9 IRAF package, utilizing
pipelines based on the NIFSEXAMPLES scripts.  Ronchi mask and arc lamp
images were used to determine the spatial and spectral geometry of the
images.  The sky images were subtracted from their neighboring
on-source images, and then these images were flat-fielded, had bad
pixels removed, and were split apart into long-slit slices.  The
spectra were then divided by the telluric calibrator (using
NFTELLURIC) after correcting for the Br$\gamma$ absorption in the
calibrator.  We made minor alterations to the Gemini pipeline to
enable propogation of the variance spectrum.  This also required
creation of our own IDL version of the NIFCUBE to make the final data
cubes.  In this process the initial pixel size of the data,
0.103\asec~in the vertical direction and 0.045\asec~in the horizontal
direction, was rebinned to 0.05\asec$\times$0.05\asec~pixels in the
final data cube for each on-source exposure.  Data cubes for the six
dithered exposures were then shifted spatially and combined with
cosmic-ray rejection using our own IDL program.  Data from the 2nd
night was spectrally shifted to the wavelength solution of the first
night before combining.  The heliocentric correction was nearly
identical on both nights (+13.25 km$\,$s$^{-1}$).  The NIFS spectrum within
a 1\asec~aperture is shown in Fig.~\ref{optirfig} and shows strong
absorption and emission lines.

We also processed the sky frames in a manner identical to the
on-source exposures.  We used the resulting data to calibrate the
spectral resolution of our observations.  Using thirteen isolated,
strong, doublet sky lines, we measured the spectral resolution of each
pixel in our final data cube.  We found significant variations in the
spectral resolution across the detector, with a median resolution of
4.36\AA~FWHM ($\lambda/\Delta\lambda = 5275$), and values ranging from
3.9 to 5.1\AA~FWHM.

A $K$-band image was made by collapsing all channels after multiplying
by (1) a 9500K black-body to correct for the shape of the telluric
spectrum, and (2) the 2MASS response curve.  Flux calibration was
obtained from 2MASS by deriving a zeropoint from images of both
sources and calibrators in our sample.  We flux calibrated our spectra
using synthetic 2MASS photometry of our telluric calibrator HIP~1123,
assuming its spectral shape resembles that of a 9500K black body.
From this flux calibration we obtained a flux calibrated spectrum of
our NGC404 NIFS data cube with units of
ergs$\,$s$^{-1}\,$cm$^{-2}\,$\AA$^{-1}$.  The flux calibration of the
image and spectrum is good to $\sim$10\%.

\subsubsection{PSF Determination}

Understanding the point spread function (PSF) is important for
interpreting the kinematic observations presented in \S4, and is even
more critical for the dynamical modeling in \S6.  We used two methods
for measuring the PSF of the calibrator: (1) we created images from
the telluric calibrator for both nights to estimate the PSF, and (2)
we compared the final NIFS $K$-band images of the NGC404 nucleus to
HST F814W data.  

We initially tried modeling the PSF as a double Gaussian \citep[as
in][]{krajnovic09}.  However, double Gaussians were a very poor fit to
the wings of the PSF in the telluric calibrator images.  Using an
inner Gaussian $+$ outer Moffat profile ($\Sigma(r) =
\Sigma_0/[(1+(r/r_d)^2]^{4.765}$) adequately described the telluric
calibrator PSF.  The best fit on both nights was a PSF with an inner
Gaussian FWHM of $0\farcs12$, and an outer Moffat profile $r_d$ of
$0\farcs95$, each containing about half the light.  The two nights
PSFs were very similar ($<10$\% differences) in both shape and fraction
of light in each component.  Unfortunately, our telluric images were
not well-dithered and thus the central PSF is not well sampled.

We further refined the PSF by fitting a convolved version of the F814W
band HST image to the NIFS $K$-band image.  
%Added Resub
Although HST/NIC2 and NIC3 F160W data were also available, the larger
pixel scale and the limited resolution of HST in the NIR
(FWHM$\sim$0.14\asec) make the F814W data a better image to use as a
PSF reference.
Due to the dust around the nucleus, we fit only the western,
apparently dust-free, half of the nucleus (Fig.~\ref{colorfig}).
Also, we find in \S3.1 that the NIFS image contains compact
non-stellar emission from hot dust at the center; we therefore correct
the $K$-band image for this dust correction before determination of
the PSF.  Finally, because the profile of the galaxy is much shallower
than the PSF, we fix the functional form of the outer profile to that
derived for the telluric star, but allow the inner PSF width and
fraction of light in each component to vary.  The best-fitting PSF has
a core with FWHM$\sim0\farcs09$ (assuming an HST resolution of
$0\farcs065$ based on TinyTim models; \citet{krist95}) containing 51\%
of the light, with the outer Moffat (with $r_d=0\farcs95$) function
containing 49\% of the light.

This PSF suffers from a number of uncertainties.  Most notably, the
presence of a bluer population at the center of the cluster implied by
Fig.~\ref{colorfig} may make the F814W image more centrally
concentrated relative to the dust-corrected $K$-band image.  Due to
our fitting the NIFS PSF core to a convolution of the F814W image,
this effect could lead to an over-estimate in the core FWHM.

\subsubsection{Stellar Kinematics Determination}

Derivation of the stellar kinematics was done as described in
\citet{seth08b} with a couple of additions.  In brief, we first
spatially binned the data to a target signal-to-noise (S/N) of 25
using the Voronoi tesselation method described in
\citet{cappellari03}.  The S/N of the data was as high as 175 in the
center, and data within $\sim0\farcs7$ was left unbinned.  Next, we
determined the stellar line-of-sight velocity distribution (LOSVD)
using the penalized pixel-fitting of \citet{cappellari04} to derive
the velocity, dispersion, $h3$, and $h4$ components in the wavelength
region from 22850\AA~to 23900\AA.  We used high-resolution templates
from \citet{wallace96} which has supergiant, giant and main sequence
stars between spectral types G and M; we used eight of these stars
with the most complete wavelength coverage.  These templates were
convolved to the instrumental resolution measured in each pixel from
the sky lines before fitting the LOSVD.  Nearly identical kinematic
results and fit qualities were obtained using the larger GNIRS library
\citep{winge08} as templates.  We performed LOSVD fits over different
ranges of wavelengths and with different subsets of templates and
achieved consistent results, suggesting template mismatch is not
a significant issue.  The higher spectral resolution and more accurate
velocity zeropoints made the \citet{wallace96} better than the
\citet{winge08} templates for deriving the kinematics.  Errors on the
LOSVD were calculated by taking the standard deviation of values
derived from 100 Monte Carlo simulations in which appropriate levels
of Gaussian random noise was added to each spectral pixel before
remeasuring the LOSVD.  Errors on the radial velocities ranged from
0.5 to 6~km$\,$s$^{-1}$ depending primarily on the S/N.

\subsection{Optical Spectroscopy}

Optical spectroscopy of the NGC~404 cluster was obtained (along with
spectra of other nearby NSCs) with the blue channel spectrograph (BCS)
on the MMT 6.5m telescope on the night of 19 Jan 2009.  We used the
500 line grating and a 1 arcsecond slit to obtain a resolution of
$\sim$3.6\AA~FWHM between 3680\AA~and 6815\AA~with a pixel size of
$0\farcs3$ spatially and 1.17\AA~spectrally.  Two 1200s exposures were
taken, with one being affected by clouds.  Due to variations in the
slit width, flat-fielding of the data was accomplished by carefully
constructing a composite flat field using the pattern from quartz
flats taken just after the exposure with pixel-to-pixel information
drawn from flats throughout the night.  Wavelength calibration was
done using arcs taken just after the science exposures.  Spectral
extraction was accomplished using the standard {\tt IRAF DOSLIT}
routine using optimal extraction over the central $2\farcs3$ of the slit
while sky regions were selected to avoid regions of emission near the
nucleus.  Guide camera seeing was around $0\farcs8$, however within the
instrument it appeared to be significantly larger ($\sim1\farcs7$).
Three flux calibrators, Gliese~191B2B, HZ44, and VMa2
\citep{bessell99} were observed on the same night with the same
observing set up.  Flux calibration was achieved by combining
information from all three flux calibrators using {\tt SENSFUNC}.
Final residuals between the fitted sensitivity function and the 3
calibrators were $<$0.03 mags between 3680\AA~and 6200\AA.  We note
that 2nd order contamination was present beyond 6200\AA.
%; we used only
%the reddish VMa2 at this wavelength for the flux calibration and
%NGC~404 shows some contamination from the 2nd order UV spectrum at
%these wavelengths.  
One of the NGC~404 exposures and 2/3 calibrator exposures were
affected by clouds.  However, by scaling to the clear exposures, the
absolute calibration of the spectra appears to be quite good;
comparison of emission line fluxes to those in \citet{ho97a} shows
agreement to within 5\% through a similar size aperture.  The S/N per
pixel of the final combined spectrum ranged from 100 at 3700\AA~to 250
around 5000\AA.  The extracted nuclear spectrum is shown in
Fig.~\ref{optirfig}.

\subsection{HST data}

\begin{figure}
\plotone{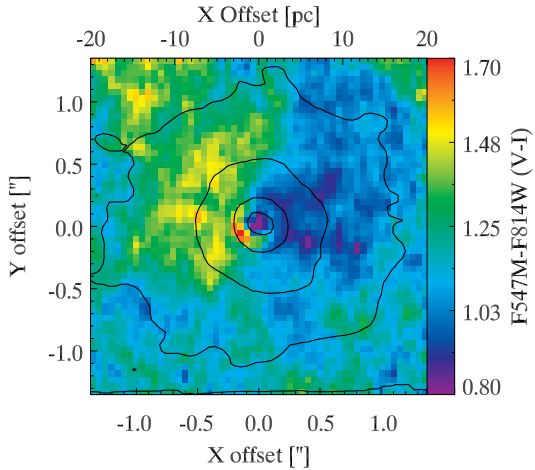}
\caption{The HST WFPC2 $F547M-F814W$ (approximately $V-I$) color map
of the nuclear region.  Contours show the NIFS continuum emission at
$\mu_K$ of 11.2, 12.2, 13.2, 14.2 and 15.2 mag/arcsec$^2$.  The image
is centered on the dust-corrected peak of the continuum light with
North up and East to the left.  Redder regions result from dust
extinction.  The western half of the nucleus appears to have little
internal extinction (see \S5.2).}
\label{colorfig}
\end{figure}

A wide variety of HST data are available for the nucleus of NGC~404
taken with the FOS, WFPC2, NICMOS and ACS HRC and WFC.  In this paper
we will use WFPC2 data in the F547M, F656N (PID: 6871), F555W, and
F814W (PID: 5999) filters, ACS WFC data in F814W filter (PID: 9293)
and NICMOS NIC2 and NIC3 data in the F160W filter (PID: 7330 \&
7919).  Data were downloaded from the HST archive and the Hubble
Legacy Archive (HLA).  The WFPC2 data all had the nucleus in the PC
chip and were unsaturated, while the ACS F814W data is saturated at
the center but covers the galaxy to larger radii than the WFPC2 data.
Astrometric alignment of this data and the NIFS data was tied to the
F814W WFPC2 data.  The centroided position of the nucleus was used to
align each image to the F814W data.  Although dust clearly affects the
area around the nucleus, the F547M-F814W color map
(Fig.~\ref{colorfig}) and the alignment of sources in the NIFS
Br$\gamma$ map with the HST H$\alpha$ image (see \S3.2) suggests that
the very center of the galaxy does not suffer from significant
internal dust extinction.

\section{Morphology}

\begin{figure*}
\plottwo{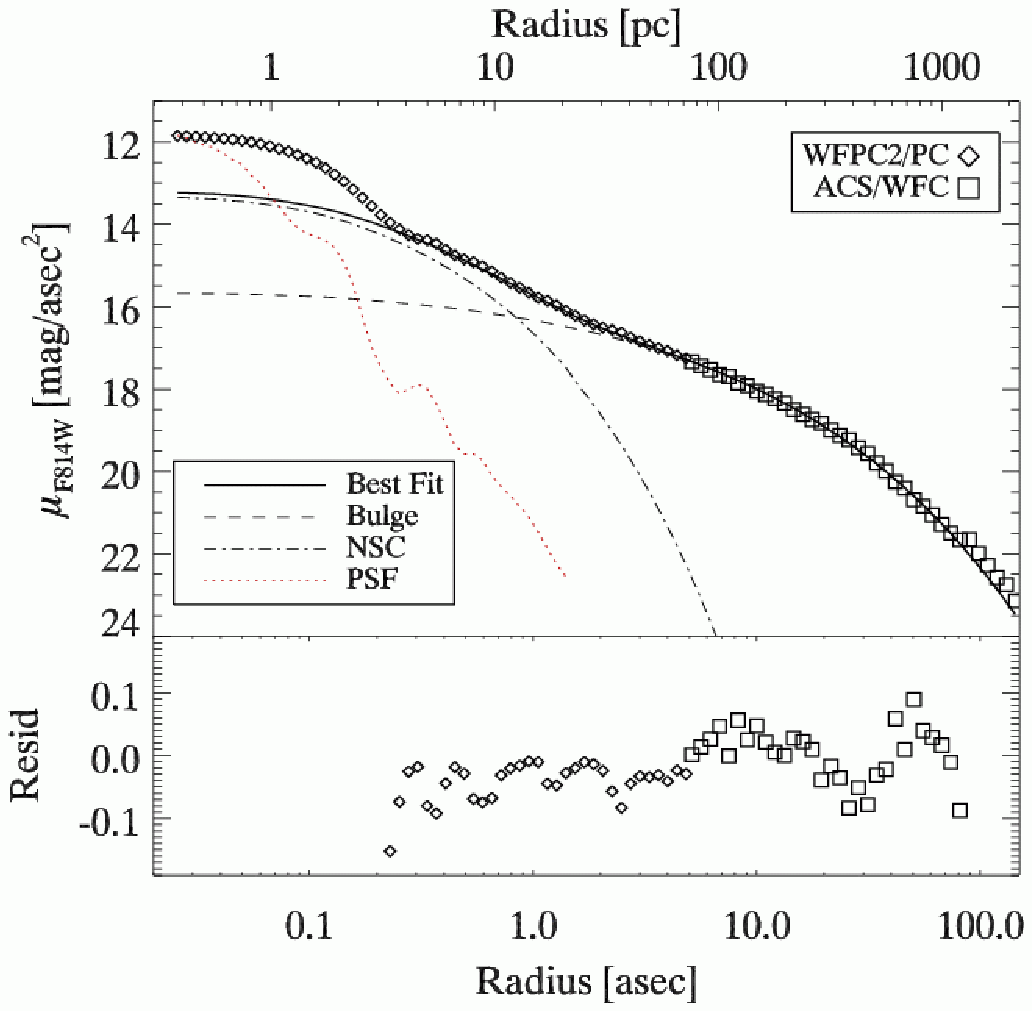}{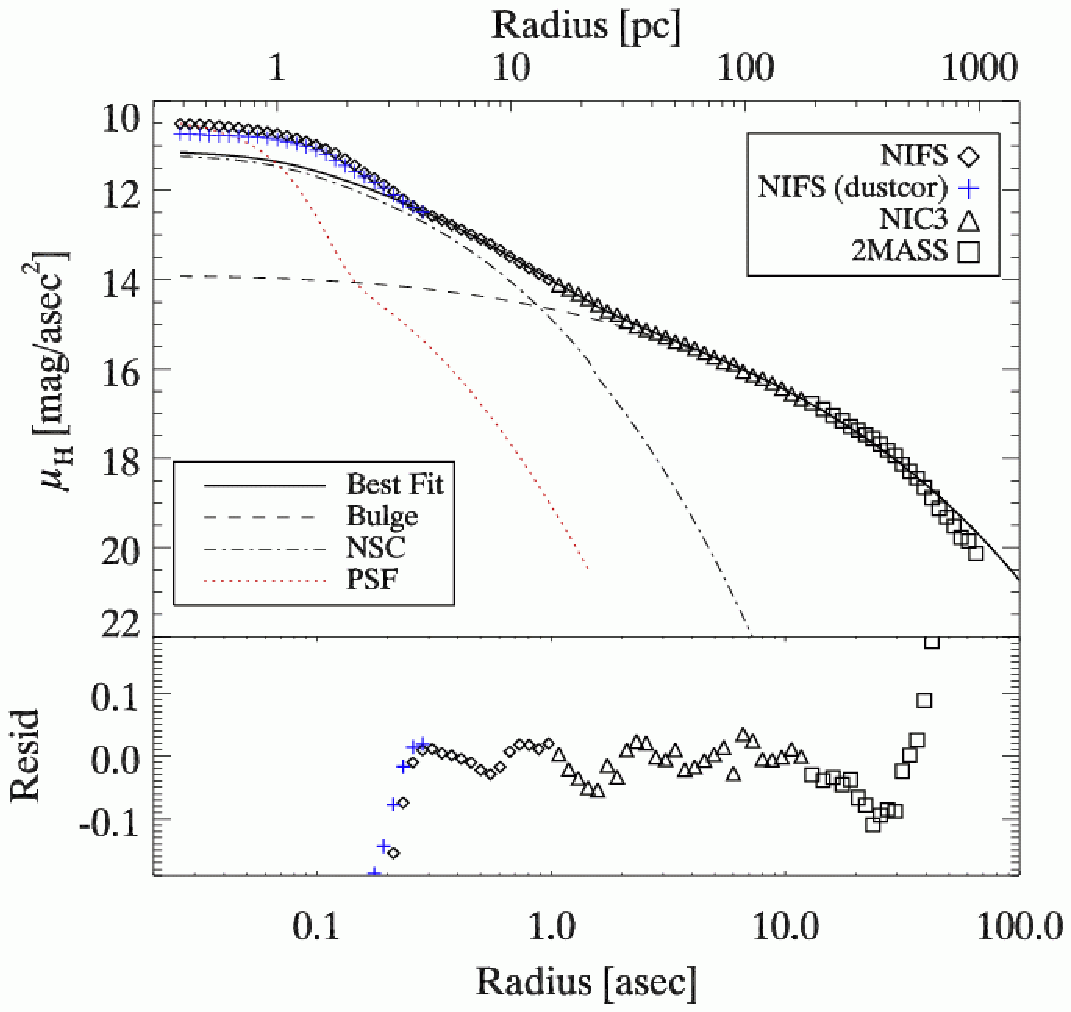}
\caption{Surface brightness profiles of NGC~404 created from HST,
Gemini/NIFS, and 2MASS data.  In both panels, the observed surface
brightness profiles are plotted with symbols, the solid line shows the
best-fit two-S\'ersic model, the dashed line the outer (bulge)
component, and the dot-dashed line the inner (NSC) component.  The PSF
is shown as a dotted red line in both plots.  Residuals disappear off
the plot at small radii to the unmodeled extra-light component.  {\it
Left --} the F814W profile constructed from both WFPC2/PC (diamonds)
and ACS/WFC (squares) data.  This data is the most reliable data
available at large radii.  {\it Right --} the NIR profile constructed
from three images: NIFS $K$-band (diamonds), HST NICMOS/NIC3
(triangles) and 2MASS large galaxy atlas data
\citep[squares][]{jarrett03}.  Blue diamonds show a correction for hot
dust emission discussed in \S3.1.}
\label{sbfig}
\end{figure*}

To determine the extent and luminosity of the nuclear star cluster in
NGC~404, it is necessary to fit the relatively steep surface
brightness profile of the inner part of the galaxy.  Using S\'ersic
fits to $I$ and $H/K$-band data, we find evidence that a nuclear star
cluster dominates the central 1\asec~(15~pc) of the galaxy, and that
an additional emission component appears to be present within the
central few parsecs (see Fig.~\ref{sbfig}).

Before describing our results on the central parts of the galaxy, we
summarize previous studies of the large scale light-profile of
NGC~404.  Using ground-based $V$-band data out to nearly 300\asec,
\citet{baggett98} decompose the SB profile of NGC404 into a $r^{1/4}$
bulge and exponential disk.  They find a bulge effective radius of
64\asec~and a disk with scale length of 130\asec.  The bulge
dominates the light out to $\sim$150\asec and the bulge-to-total ratio
is $\sim$0.7.  \citet{tikhonov03} found that the number counts of red
giant branch stars between 100-500\asec~(1.5-7.5~kpc) are also well
explained by an exponential disk with a similar, $\sim$106\asec~scale
length.

We used archival HST data, our NIFS $K$-band image, and the 2MASS
Large Galaxy Atlas \citep{jarrett03} to construct the two surface
brightness profiles shown in Fig.~\ref{sbfig}.  Surface brightness
measurements were obtained on all images using the {\tt ellipse}
program in IRAF.  The F814W-band ($I$) profile combines HST WFPC2 data
from the PC chip at radii below 5\asec~with ACS/WFC data at larger
radii (the ACS image was saturated at smaller radii).  Diffraction
spikes from $\beta\,$And (7' to the SE) and bright foreground stars
and background galaxies were masked.  Dust is obviously present near
the center and this was masked out as well as possible using a
$F555W-F814W$ color map at radii between $0\farcs3$ and 5\asec~from
the center.  Comparing this profile with the NIR light profile
suggests that some dust extinction remains within the inner
1\asec. The NIR profile combines the NIFS $K$-band image (at $r \leq
1$\asec), with the F160W-band ($H$) NICMOS/NIC3 image ($1 < r \leq
5$\asec) and $H$ band data from the 2MASS Large Galaxy Atlas (at $r >
5$\asec).  The NIFS data was scaled to match the $H$-band zeropoint;
no scaling was needed to match the 2MASS and NICMOS data.  The final
F814W and NIR surface brightness profiles are shown in
Fig.~\ref{sbfig}.

We first fit the outer part of both profiles to a S\'ersic profile
\citep[see][for a complete discussion of S\'ersic profiles]{graham05}.
During our fitting, the S\'ersic profiles were convolved with the PSF.
The 2MASS data was only usable to 70\asec~due to low S/N, and even
within this radius the sky subtraction is suspect due to nearby
$\beta$And.  The F814W data is more reliable at these larger radii,
however we fit only the measurements within 80\asec~due to uncertainty
in the sky background and possible contribution from a disk at larger
radii.  At radii beyond 5\asec, the profiles are well fit by similar
S\'ersic profiles with $n \sim 2.5$ and effective (half-light) radius
of $\sim$43\asec~(640~pc).  Exact values for all fits are given in
Table~\ref{sersictab}.  We identify this outer component as the bulge
given the results of \citet{baggett98}, who show the bulge component
dominates at $r \lesssim 150$\asec.  It was not possible to fit either
profile down to smaller radii with a single S\'ersic.

Following \citet{graham03,cote06}, we identify the excess light over the bulge
at small radii to be a NSC.  We discuss the nature of this component
more in \S7.1.  In both profiles, the NSC component is brighter than
the bulge at $r < 1$\asec.  We fit this component with an additional
S\'ersic component \citep[following][]{graham09} and find it has an
effective radius of $\sim0\farcs7$ (10~pc).  The NSC S\'ersic $n$
value is quite sensitive to the functional form of the PSF, therefore
we prefer the F814W value of $n$=1.99 due to the better-known PSF.
These parameters are in the ranges typical of NSCs \citep{cote06,
graham09} in other galaxies. The absolute magnitude of this component
(based on the model) is $M_I, M_H=-13.64,-15.34$ within the central
3\asec, and $M_I, M_H=-13.17, -14.93$ within 1\asec.

In the inner $0\farcs2$, there is a clear excess luminosity above the
double-S\'ersic fit in both profiles.  Adding up the light in excess
of the profile we find an $M_I, M_H=-11.81, -12.61$.  Note that these
magnitudes are quite sensitive to the PSF and S\'ersic $n$ indices of
the NSC fit.  We will show below that in the NIR, a fraction of this
light appears to be emission by hot 950~K dust.  A correction for this
dust emission is shown with blue diamonds in Fig.~\ref{sbfig}.
However, dust alone cannot explain the luminosity of this excess
light, accounting for only $\sim$35\% of the excess in the NIR
profile.  Furthermore, at $I$-band, no significant dust emission would
be expected at all.  Two additional clues paint a complicated picture.
First, the color in the central $0\farcs2$ is somewhat bluer in the
HST $F547M-F814W$ images (Fig.~\ref{colorfig}) and has compact UV
emission \citep{maoz05}, both suggestive of the presence of young
stars.  Second, the stars in this area appear to be counter-rotating
relative to the NSC and galaxy (see \S4.1).  Thus the central excess
appears to result from a combination of hot dust emission, younger
stars, and perhaps AGN continuum emission in the optical and UV.

In summary, we find evidence for 3 components at the center of NGC~404.  
\begin{enumerate}
\item A bulge with S\'ersic $n \sim 2.5$ and effective radius of
640~pc that dominates the light between 1\asec~and 80\asec~(and beyond).
\item A nuclear star cluster which dominates the light in the central
arcsecond (15~pc).
\item A central excess at $r < 0\farcs2$.  This excess appears to
result from the combination of hot dust, young stars, and perhaps AGN
continuum emission.
\end{enumerate}

Within the central 1\asec~in the $I$-band (roughly the area included
in the MMT spectrum analyzed in \S5), $\sim$55\% of the light comes
from the NSC, 29\% from the underlying bulge, and 16\% 
from the central excess.  
%The bluer central excess may contribute a
%somewhat larger fraction of the light to our spectrum at
%$\sim$5000\AA.

The isophotes in the ACS F814W data at 10-80\asec have a position
angle of $\sim$80 degrees and an ellipticity of $\sim$0.08.  This
matches well with the morphology of the HI disk on a large scale;
\citet{delrio04} finds an HI inclination of $\sim$10 degrees and a
position angle of about $\sim$80 degrees for radii between
100-300\asec.  Due to dust absorption, the clearest picture of the
inner arcsecond of NGC~404 comes from the NIFS $K$-band image.  As can
be seen from the contours in Fig.~\ref{colorfig}, the position angle
at $r < 0\farcs25$ has a PA of 60-70 degrees.  The PA then abruptly
changes to 25 degrees at $r = 0\farcs3$ and gradually increases to the
large scale PA of 80 degrees by $\sim$1\asec.  The ellipticity
decreases from $\sim$0.2 near the center to 0.07 at 1\asec.  These
changes fit well with our picture of three distinct components, each
having their own distinct morphology.

\begin{deluxetable}{lcccc}
\tablewidth{0pt} 
\tablecaption{Morphological Fits \label{sersictab}}
\startdata

\tableline
\multicolumn{5}{c}{{\it F814W Profile}}\\
Component & S\'ersic $n$ &  $r_{eff}$ [\asec] & $\mu_{eff,F814W}$ [mag/\asec$^2$] & $M_I$ \\
\tableline
Bulge & 2.43 & 41.5 & 20.18 & -18.09 \\
NSC   & 1.99 & 0.74 & 16.02 & -13.64 \\
\multicolumn{4}{l}{Central Excess} & -11.81 \\

\tableline
\multicolumn{5}{c}{{\it NIR Profile}}\\
Component & S\'ersic $n$ &  $r_{eff}$ [\asec] & $\mu_{eff,H}$ [mag/\asec$^2$] & $M_H$ \\
\tableline

Bulge   & 2.71 & 44.7 & 18.82 & -19.62 \\
NSC     & 2.61 & 0.68 & 14.15 & -15.34 \\
\multicolumn{4}{l}{Central Excess} & -12.61 \\
\multicolumn{4}{l}{Central Excess (Dust Corrected)} & -12.14
\enddata 

\tablecomments{Bulge magnitudes given for $r \leq 80$\asec, and NSC
magnitudes for $r \leq 3$\asec.  Central excess magnitudes are for the
total flux above the fit in the central arcsecond.  Distance modulus
is assumed to be 27.43 giving 1\asec=14.8~pc.  Note that the
magnitudes are not corrected for foreground extinction
\citep[$A_I=0.11$, $A_H=0.03$;][]{schlegel98}.}

\end{deluxetable}

\vspace{0.1in}

\subsection{A Hot Dust Component}

\begin{figure}
\plotone{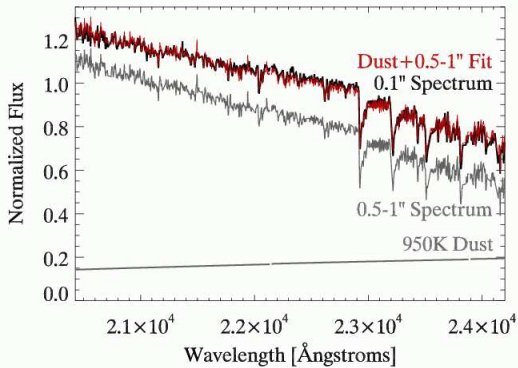}
\caption{Fit of the central NIFS spectrum of NGC~404 to a model
including emission from hot dust.  The black line shows the central
spectrum, with the overplotted line showing the best fit.  This fit
consists of a template stellar spectrum taken from an annulus of
radius $0\farcs5$-$1\farcs0$, plus a black-body of temperature 950~K
(shown in gray).  Note that emission lines were excluded from the fit
and plot.}
\label{bbfig}
\end{figure}

Within the central 0.2'', the NIFS spectra show a significant spectral
flattening and a reduction in the depths of the CO lines by about
10\%.  A comparison of the central spectrum to single stellar
populations models from \citet{maraston05} shows that its spectral
slope is flatter/redder than any stellar population models.
Furthermore, the younger stellar populations expected in the central
excess should have a somewhat bluer, not redder, continuum.  This
redder continuum suggests two possibilities (1) strong dust absorption
or (2) emission from hot dust.

We rule out dust absorption as a possibility because the required dust
absorption to flatten the spectrum is $A_V \sim 6$ magnitudes assuming
the reddening law from \citet{cardelli89}.  The $F547M-F814W$ shows that
the nucleus is actually {\it bluer} than the surrounding areas
($F547M-F814W$ $\sim$ 0.8).  Furthermore, the area to the west of the
nucleus appears unreddened with $F547W-F814W \sim 0.95$, consistent
with the expected color of the stellar population based on
spectroscopic fits (see \S5.2), while the eastern half of the nucleus
is at most 0.8 magnitudes redder in $F547M-F814W$ (approximately
$V-I$).  This suggests a maximum extinction in the region around the
nucleus of $A_V \sim 2$.

Therefore, hot dust appears to be the cause of the flattened continuum
in the central pixels.  Hot dust is a common feature in Seyfert nuclei
at NIR wavelengths, and is typically found to have temperatures of
800-1300~K \citep{alonso-herrero96, winge00, riffel09}.  Studies of
Seyfert galaxies with NICMOS by \citet{quillen00b,quillen01b} show
that the hot-dust component is frequently unresolved and variable on
months to year timescales.  The presence of hot dust components in
LINER galaxies appears to be unstudied.

We fit the contribution and temperature of this hot dust emission by
assuming that the central spectrum is made up of a nuclear star
cluster spectrum drawn from an annulus of $0\farcs5$-$1\farcs0$ plus a
pure blackbody spectrum.  
As shown in Fig.~\ref{bbfig}, the central spectrum ($r \leq
0\farcs05$) is best fit by a dust temperature of 950~K, contributing
about 20\% of the light.  The residuals between the central spectra
and model have a standard deviation of 1.4\% across the band and the
fit completely eliminates the difference in spectral slopes between
the annular and central spectrum.

Fitting spectrum in radial annuli, we find no evidence for dust
emission beyond $0\farcs3$ from the nucleus.  Assuming a uniform dust
temperature of 950~K, we calculated the contribution of dust in each
individual spaxel.  Using this, we correct the $K$-band image for the
hot dust contribution.  {\em We use the dust-corrected $K$-band image
and central position for all our subsequent analysis.}  The dust
emission accounts for only $\sim$35\% of the central excess in the
NIR, as is shown with the blue points in right panel of
Fig.~\ref{sbfig}.

The dust emission is slightly offset to the NE of the photocenter.
This offset is in the direction of the dust absorption clearly seen on
larger scales to the E of the cluster in the HST color image
(Fig.~\ref{colorfig}).  The corrected image has a photocenter offset
by about a half-pixel from the uncorrected image, while the dust
emission is offset by $\sim$1.5 pixels from the corrected photocenter.
The small shift in the photocenter reduces the asymmetry in the
velocity field, moving the photocenter to the center of the central
counter-rotation (see \S4.1).  As found in other AGN, the hot dust
component is unresolved, with a measured FWHM of $0\farcs14$
(single-Gaussian PSF fits give a FWHM of $\sim0\farcs18$ for our NIFS
data). 

The total dust emission in $K$ band has a magnitude of $K \sim 15.4$
which translates to a flux of $1.3 \times
10^{-13}$~ergs$\,$s$^{-1}\,$cm$^{-2}$ and luminosity of $1.4
\times 10^{38}$~ergs$\,$s$^{-1}$.  Assuming the hot dust emission is
the result of accretion onto a BH of $\sim 10^5$~M$_\odot$ (\S6), this
suggests $L_{bol}/L_{Edd} \gtrsim 10^{-5}$; typical LINERS have
$L_{bol}/L_{Edd} \sim 3 \times 10^{-5}$ \citep{ho04}.  The hot dust
luminosity is somewhat larger than the [\ion{O}{3}] luminosity
(assuming D=3.06~Mpc) of $2 \times 10^{37}$~ergs$\,$s$^{-1}$
\citep{ho97a} and the possible hard X-ray component with 2-10 keV
luminosity of $\sim3 \times 10^{37}$~ergs$\,$s$^{-1}$ noted by
\citet{terashima02}.  The 1.6 $\mu$m dust emission, [\ion{O}{3}], and
hard X-ray luminosities of Seyfert galaxies are discussed in detail by
\citet{quillen01b}.  Although the luminosities presented here are much
lower than those in their sample (as would be expected for a system
with a lower accretion rate), the relative luminosities in these three
bands fall within the scatter of their data suggesting a comparable
SED.

\begin{figure}
\plotone{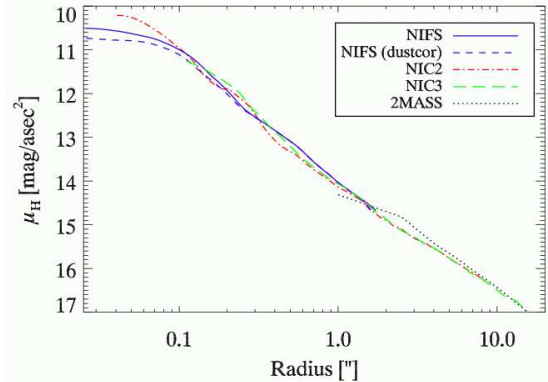}
\caption{The surface brightness profile of the NICMOS F160W
observations (NIC2 in red, NIC3 in orange), 2MASS $H$-band (dashed),
and NIFS $K$-band observations scaled to match the F160W NICMOS
observations at $r > 1$\asec.  The dashed line shows the NIFS
observations after correction for dust emission.}
\label{niccompfig}
\end{figure}

\begin{figure*}
\epsscale{1.2}
\plotone{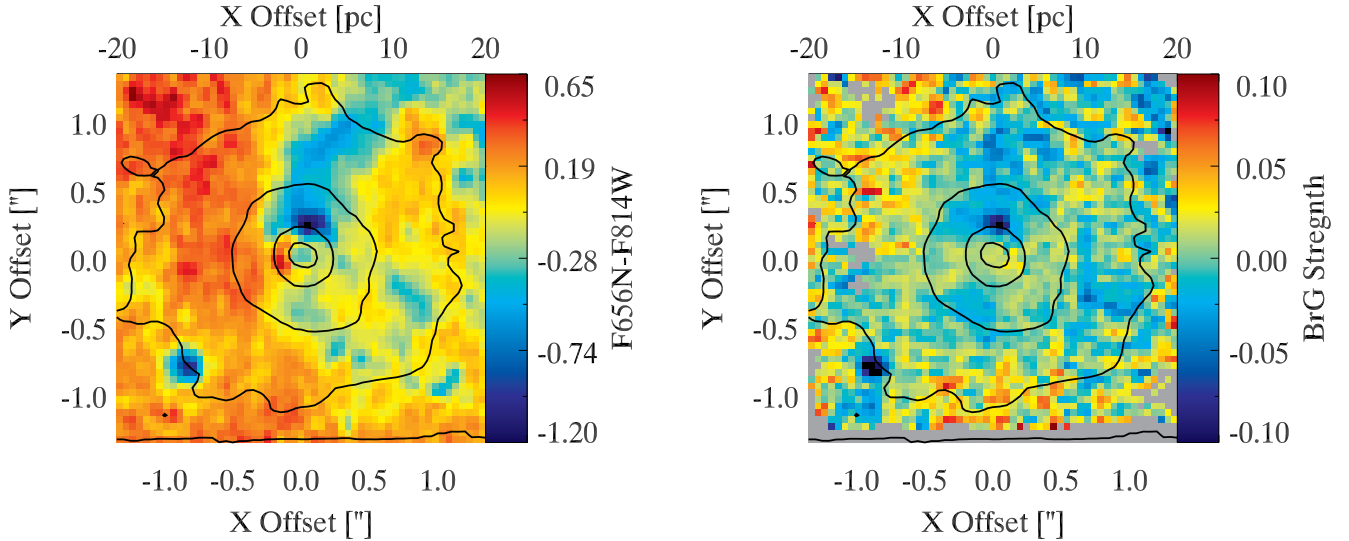}
\caption{{\it Left --} a WFPC2 F656N-F814W map of the nuclear region.
Negative values are areas with strong H$\alpha$ emission.  {\it Right --} Br$\gamma$ emission line map.  Negative values mean strong Br$\gamma$ emission, while positive values result from Br$\gamma$ absorption. Contours and orientation are identical to those in Fig.~\ref{colorfig}.}
\label{halphabrgfig}
\end{figure*}

We also find tentative evidence for variability in the dust emission
between our NIFS images taken in 2008 and HST NICMOS images taken in
1998.  This variability may be expected given the variability in the
nuclear UV flux reported by \citet{maoz05}.  They find that the flux
at 2500\AA~decreasing by a factor of $\sim$3 between 1994 and 2002.
However, this comparison is based on observations by three different
cameras, the earliest of which was an $0\farcs86$ aperture spectrum,
thus spatial resolution of the UV variability is not possible.

Figure \ref{niccompfig} shows a comparison of the surface brightness
profiles of all the NIR data, with the NIFS $K$-band image scaled to
fit the SB profile of the NICMOS F160W images at $r > 1$\asec.  Within
the central $0\farcs2$, the scaled NIFS luminosity is clearly fainter
than the NIC2 image.  This change in brightness could be explained by
a bluer stellar population within the central excess, but the effect
is much larger than expected -- the $F547M-F814W$ image shows a change
of $\Delta(F547M-F814W) \sim 0.1$~mags between the center and the NSC
at larger radii, implying a $\Delta(H-K) < 0.1$.  However, the
observed $\Delta(H-K) = 0.4$ after attempting to correct both
the NIC2 image and NIFS dust-subtracted image for the effects of the
PSF.  Using the original NIFS image (without dust subtraction), we
still get $\Delta(H-K) = 0.3$.  We suggest that this large
$\Delta(H-K)$ may result from variable dust emission, with the NIC2
observations being taken at an earlier epoch when the dust-emission
was more prominent.  Multi-epoch HST observations of the nucleus in
the UV and NIR using the same instrument will be extremely useful for
verifying a possible variable AGN contribution and constraining its
SED.

\begin{figure*}
\epsscale{1.2}
\plotone{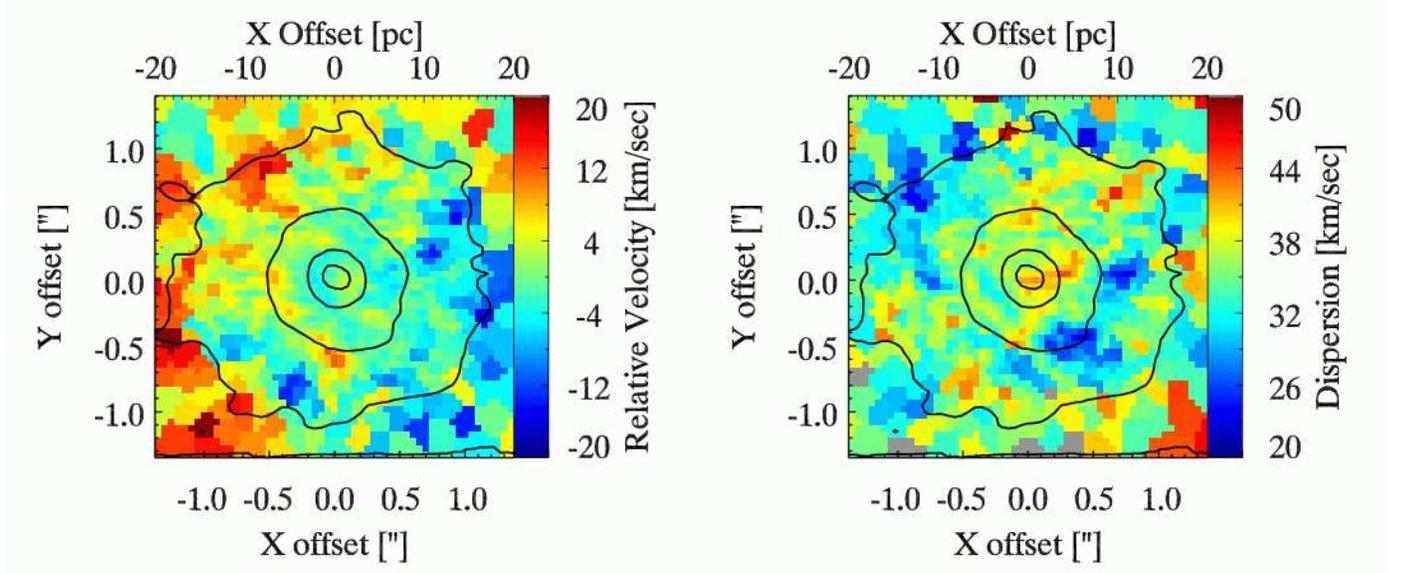}
\caption{Velocity and dispersion of the stellar component derived from
the CO bandhead.  Contours show contours of the $K$-band image with
contours separated by a magnitude in surface brightness (as in
Fig.~\ref{colorfig}).  Radial Velocities are shown relative to the
central velocity of -58.9 km$\,$s$^{-1}$.}
\label{abs2dfig}
\end{figure*}

\begin{figure}
\epsscale{1.1}
\plotone{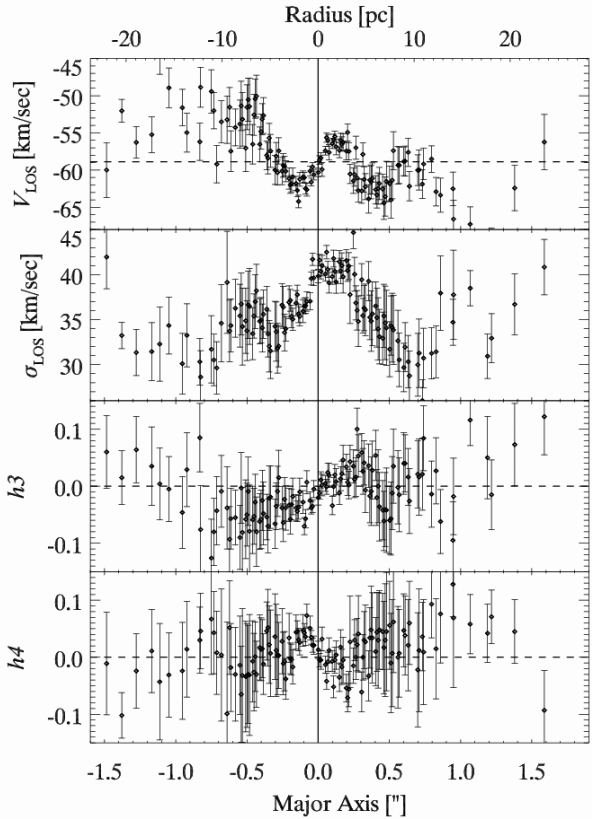}
\caption{The LOSVD along the major axis of the innermost component at
PA=80$^\circ$.  Panels show the radial velocity, LOS dispersion, $h3$ and
$h4$ components of the LOSVD (from top to bottom).  Bin centers
falling within $0\farcs1$ of the major axis are plotted.}
\label{majoraxisfig}
\end{figure}

\subsection{NIR Emission Line Excitation}

In this section we discuss the emission lines observed in the
Gemini/NIFS $K$-band spectrum.  We find that they imply that the
emission originates in dense molecular gas with line ratios typical of
those observed in known AGN.

To obtain estimates for line emission in the NGC~404 nucleus, we summed
together flux-calibrated integrated spectra from our NIFS data with
radii ranging between 1 and 25 pixels.  We then used both the
\citet{wallace96} and \citet{winge08} templates to fit the stellar
continuum from 20200\AA~to~24100\AA~excluding the parts of the
spectrum near known emission lines.  Subtracting this continuum off
left us with a pure emission spectrum.  We measured numerous H$_2$
emission lines as well as Br$\gamma$ emission.  The only observable
coronal line commonly seen in AGN is [CaVIII], which falls directly on
a CO bandhead; an $3-\sigma$ upper limit of $3.6 \times
10^{-17}$~ergs$\,$s$^{-1}\,$cm$^{-2}$ was placed on this line.

Detected H$_2$ lines (and their vacuum wavelengths) included 1-0~S(2)
(20337.6\AA), 2-1~S(3) (20734.7\AA), 1-0~S(1) (21218.3\AA), 2-1~S(2)
(21542.1\AA), 1-0~S(0) (22232.9\AA), 2-1~S(1) (22477.2\AA) and
1-0~Q(1) (24065.9\AA).  Line ratios of these lines can be used to
determine the excitation mechanisms of the gas.  In particular, the
ratio of the 2-1~S(1) and 1-0~S(1) lines is between 0.11 and 0.16 with
smaller values at larger radii.  These values are indicative of
thermal excitation from shocks or X-ray photons in a dense cloud
($n\gtrsim10^{4}$~cm$^{-3}$), while fluorescent emission in a low density
medium has a higher 2-1~S(1)/1-0~S(1) of $\sim$0.55
\citep[e.g.][]{mouri94}.  The observed ratio is typical of those
observed in other AGN \citep[e.g.][]{reunanen02,riffel08,storchi09}.

The rotational and vibrational temperatures can be estimated from
ratios of H$_2$ lines; using the relation from \citet{reunanen02}
$T_{vib} = 5600 / ln(F(1-0 S(1))/F(2-1 S(1))*1.355)$, we find $T_{vib}
= 2340 \pm 160$~K within an 1\asec~radius aperture.  Similarly, the
rotational temperature can be determined using the relation $T_{rot} =
-1113 / ln(F(1-0 S(2))/F(1-0 S(0))*0.323)$ \citep{reunanen02} which
gives a somewhat lower temperature of $T_{rot} = 1640 \pm 360$~K
within a 1\asec~aperture.  The higher vibrational temperature may be
indicative of some contribution of fluorescence
\citep[e.g.][]{riffel08}.  

Unfortunately the Br$\gamma$ emission line flux can not be accurately
estimated due to uncorrected telluric absorption in the vicinity of
the Br$\gamma$ line.  However, the distribution of Br$\gamma$ flux is
much patchier than the H$_2$ (see Fig.~\ref{halphabrgfig}) suggesting
different sites for the two types of emission.  Note that in
starbursting galaxies, the Br$\gamma$ is stronger than the H$_2$
1-0~S(1) line \citep{rodriguez04}, a possibility which is clearly
excluded by our data (see Fig.~\ref{optirfig}).  This again is
consistent with emission originating due to thermal excitation in
dense gas as suggested by the H$_2$ 2-1~S(1)/1-0~S(1) ratio.  

The distribution of Br$\gamma$ and H$\alpha$ visible in
Fig.~\ref{halphabrgfig} emission is primarily concentrated in two
clumps, one just to the N of the nucleus and one $\sim$1\asec~to the
SE.  These are likely the sites of the most recent star formation in
the nucleus.  In \S5 we find a total young stellar mass at ages
$\lesssim 10$~Myr of $\sim10^4$~M$_\odot$.  The central excess has
bluer optical colors and UV emission but lacks strong H$\alpha$ or
Br$\gamma$ emission, suggesting the young stars there have an age of
$\gtrsim10$~Myr \citep{gogarten09}.

\section{Kinematics}

Our NIFS data enable us to derive kinematics for both the absorption
lines, using the CO bandheads from 22900-24000\AA~and the strongest of
the H$_2$ lines at 21218\AA.  

\subsection{Stellar Kinematics}

\begin{figure}
\epsscale{1.2}
\plotone{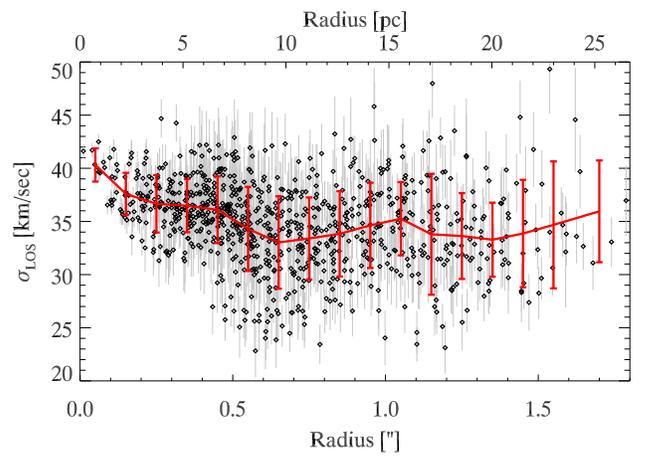}
\caption{The LOS dispersion values as a function of radius for all
data points with dispersion errors less than 10~km$\,$s$^{-1}$.  Error bars on
individual points can be seen in light gray.  The thick line shows
a binned average and standard deviation as a function of radius. }
\label{radialdispfig}
\end{figure}

\begin{figure}
\plotone{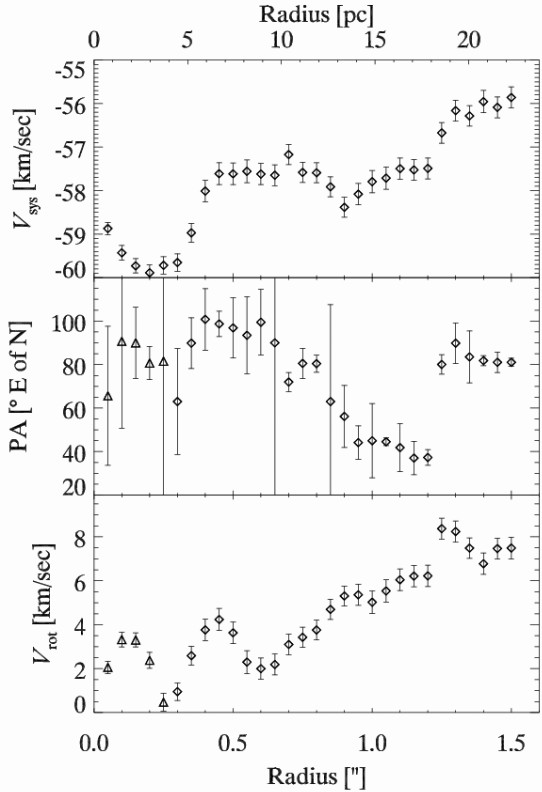}
\caption{Results of kinemetric analysis of the stellar velocity field as function of radius.  {\it Top --} the systemic velocity, {\it middle --} the position angle, {\it bottom --} the amplitude of rotation.  Triangles in bottom two panels indicate {\it counter-rotation}.}
\label{abskinfig}
\end{figure}

The velocity and dispersion across the NIFS FOV is shown in
Fig.~\ref{abs2dfig}.  At radii of $\sim$1\asec, we see a blue-shift to
the W and red-shift to the E, rotation that is in the same direction
as the HI rotation seen at larger radii \citep{delrio04}.  At $r \lesssim
0\farcs2$, there appears to be modest counter-rotation.  An
alternative view of the velocity field is seen in
Fig.~\ref{majoraxisfig} which shows the LOSVD profile along a
PA=80$^\circ$.  The error bars in the velocity range from
$<$1~km$\,$s$^{-1}$ near the center to $\sim$5~km$\,$s$^{-1}$ at large
radii.  The counter-rotation seen in the central excess (at $r <
0\farcs2$) is clearly significant.  However, given the nearly face-on
orientation of the larger scale of the galaxy, this counter-rotation
does not necessarily imply a large ($>$90$^\circ$) difference in the
angle of the rotation axes of the two stellar components.

To quantify the position angle (PA) and amplitude of the rotation as
a function of radius, we used the kinemetry program of
\citet{krajnovic06}.  This kinemetry program uses a generalized
version of ellipse fitting to analyze and quantify the velocity map or
other higher order moments of the LOSVD.  We present the results of
the kinemetry analysis of the velocity maps in Fig.~\ref{abskinfig},
showing the systemic velocity, PA, and rotation speed as a
function of radius.  We limited the axial ratio to between 0.8 and
0.95 (consistent with our ellipse fits to the $K$ band image), and
fixed the center to the photocenter of the dust-corrected $K$ band
image.  We find an amplitude for the central counter-rotation of
3.3~km$\,$s$^{-1}$ (shown in triangles in Fig.~\ref{abskinfig}), while
the rotation at larger radii has a maximum rotation amplitude of 8.4
km$\,$s$^{-1}$ at $1\farcs2$.  The PA of the counter-rotation and
rotation at larger radii are quite similar, with values of
80-90$^\circ$ (E of N).
Our observed rotation is at a similar PA to the HI gas, which has a
rotation axis of 70-90$^\circ$ from 100-300\asec~\citep{delrio04}.
The maximum observed rotation velocity of the HI gas is
$\sim$45~km$\,$s$^{-1}$.  Recent observations of the stellar kinematics along
the major axis by \citet{bouchard10} suggest that increasing
counter-rotation relative to the HI is seen between radii of
1-20\asec, with rotation in the sense of the HI again taking over at
larger radii.  Clearly, large-scale IFU kinematics of this galaxy
would be very interesting.

A radial profile of the stellar dispersion is shown in
Fig.~\ref{radialdispfig} for bins with uncertainties in the dispersion
of $<$10~km$\,$s$^{-1}$.  While the scatter of values is large at most radii,
the velocity dispersion clearly declines from $\sim$40~km$\,$s$^{-1}$
at the center to a mean value of $\sim$33~km$\,$s$^{-1}$ at $0\farcs6$.  Our
dispersion values are consistent with the 40$\pm$3~km$\,$s$^{-1}$ determined
by \citet{barth02} for the NGC~404 center using the calcium triplet.
Dynamical modeling based on the stellar kinematics is detailed in
\S6.1.

\begin{figure*}
\epsscale{1.2}
\plotone{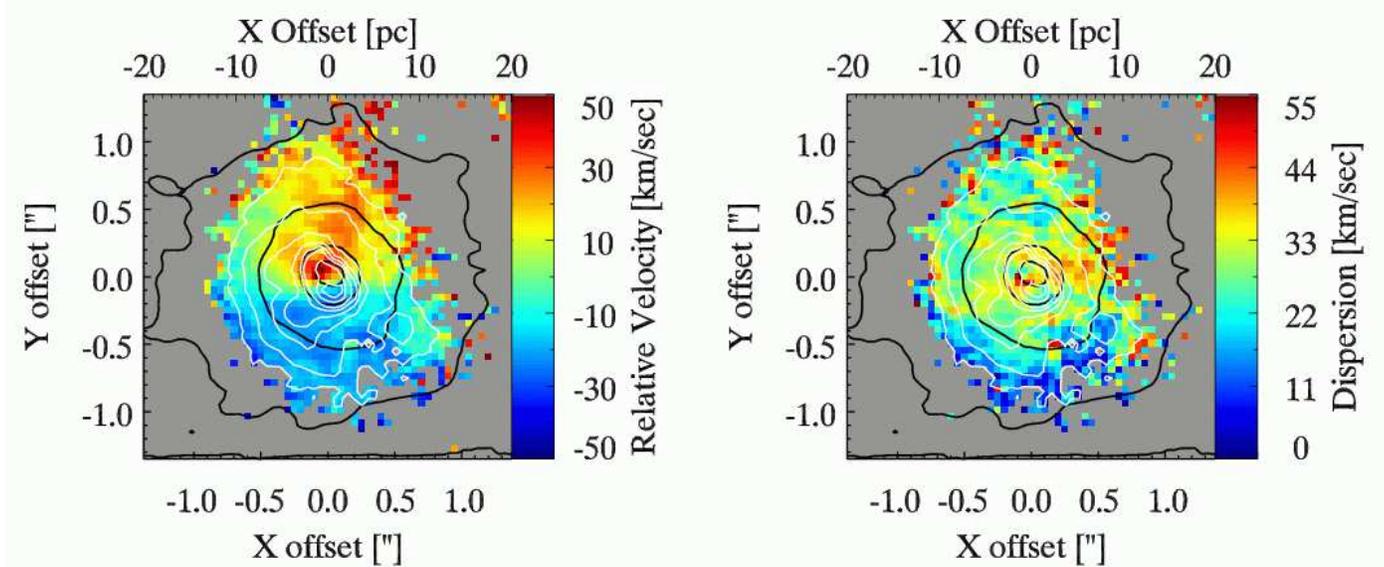}
\caption{Velocity and dispersion of the molecular hydrogen data
determined from the 1-0~S(1) line at 21218\AA.  Black contours
indicate the surface brightness of the $K$-band image with contours
separated by a magnitude in surface brightness.  White contours show
the H$_2$ total intensity 5, 10, 20, 30, 40, 50, 70,and 90\% of the
peak.  Radial velocities are shown relative to the central stellar
radial velocity (-58.9 km$\,$s$^{-1}$).}
\label{emveldispfig}
\end{figure*}

\begin{figure}
\plotone{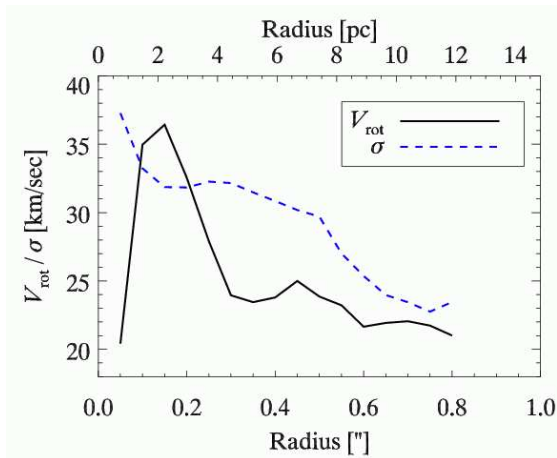}
\caption{The rotation velocity (black line) and dispersion (blue
dashed line) derived from H$_2$ 21218\AA~line using the kinemetry
program of \citet{krajnovic06}.}
\label{emradialfig}
\end{figure}

\subsection{H$_2$ Gas Kinematics and Morphology}

The H$_2$ gas kinematics and morphology were derived by fitting the
21218\AA~1-0~S(1) emission line, the strongest of the H$_2$ emission
lines in the $K$-band.  In each spaxel, we fit this line with a
Gaussian to derive a flux, radial velocity, and velocity dispersion
(after correction for the instrumental resolution).  The results of
these fits are shown in Fig.~\ref{emveldispfig}. The physical
condition of this gas is discussed more in \S3.2.  We note that the
mass of luminous excited H$_2$ gas is very small ($<1$~M$_\odot$) and
although it likely tracers a larger mass of colder molecular gas, this gas is
not expected to be dynamically significant in the nucleus
\citep{wiklind90}.

The H$_2$ emission peaks slightly to the SW of the continuum emission,
$\sim0\farcs07$ from the centroid of the dust emission corrected
continuum (see white contours in Fig.~\ref{emveldispfig}).  The FWHM
of the distribution is $\sim0\farcs4$ (6~pc).  At $r \lesssim
0\farcs3$, it is elongated ($q \sim 0.6$) with PA$\sim$25$^\circ$.

The H$_2$ gas velocity field provides a significant contrast with the
stellar velocity field.  It shows a clear rotation signature at a
position angle nearly orthogonal to the stellar rotation.  Using the
kinemetric analysis program \citep{krajnovic06}, we find a kinematic
PA of $\sim$15$^\circ$ out to $0\farcs2$, which then twists to PA $\sim
-10$ at larger radii.  A radial profile of the rotation velocity and
dispersion are shown in Fig.~\ref{emradialfig}.  The rotation peaks at
$r \sim 0\farcs15$ (2.2~pc) with an amplitude of 36~km$\,$s$^{-1}$. Beyond
$0\farcs3$, the rotation flattens out to an amplitude of 20-25~km$\,$s$^{-1}$.
The central dispersion ($r = 0\farcs05$) is 39~km$\,$s$^{-1}$, which likely
includes unresolved rotation.  At $r=0\farcs1$ it drops to 33~km$\,$s$^{-1}$,
then slowly declines to $\sim$25~km$\,$s$^{-1}$ at larger radii.  Dynamical
modeling of the gas kinematics are presented in \S6.2.

The axial ratio of the intensity distribution and our gas dynamical
model suggest an inclination of $\sim$30$^\circ$ at
$r < 0\farcs3$, with smaller inclinations at larger radii.  Assuming
this geometry, the H$_2$ disk appears to be rotation dominated at all
radii with $V_{\rm rot}/\sigma \sim 2$ near the center.
H$_2$ emission line disks are commonly observed in nearby AGN
\citep{neumayer07, riffel08, hicks09}, although the size of these
disks are typically much larger than in NGC~404.  For instance, in the
\citet{hicks09} sample of 11 nearby AGN, they find H$_2$ disks traced
by the 21218\AA~line with half-width half-maximum values ranging from
7-77~pc (vs. 3~pc in NGC~404).  Most of these disks have $V_{\rm
rot}/\sigma$ values of $\sim$1, although a couple have $V_{\rm
rot}/\sigma > 2$ at larger radii.

\section{Stellar Populations}

In this section we use the optical spectrum of the NGC~404 nuclear
star cluster to constrain its formation history.  We first use the
optical emission lines to constrain the reddening and derive a maximum
current SFR of $1.0 \times 10^{-3}$~M$_\odot$/yr.  Fits to the
absorption line spectrum show that the nucleus has a range of stellar
ages, with stars $\sim$1~Gyr in age contributing more than half the
light of the spectrum. Next, we consider the range in mass-to-light
ratios ($M/L$s) allowed by the stellar populations fits.  Finally, we
show that the bulge has a distinctly older stellar population than the
nucleus.

\vspace{0.2in}

\subsection{Optical Emission Line Information}

Before discussing our fits to the optical absorption line spectrum, we
first use the optical emission lines to estimate the reddening,
current metallicity and star formation rate (SFR).  We note again that
our line fluxes and ratios agree very well with those in \citet{ho97a}.

After correcting for the underlying stellar population using our
population fits described below, we find a Balmer decrement
($F(H\alpha)/F(H\beta)$) of 3.59.  For an AGN the expected Balmer
decrement is 3.1 while for an HII region it 2.85 \citep{osterbrock89}.
Given the ambiguous nature of the nuclear emission, this suggests
extinctions of $A_V = 0.43-0.73$.  As noted in \S3.1, the
$F547W-F814W$ color-map indicates extinctions of $A_V \lesssim 2$
covering about half the nucleus, thus these Balmer decrement
extinction values seem reasonable.  We note that the patchy
distribution of the H$\alpha$ emission (see Fig.~\ref{halphabrgfig})
is significantly different from the stellar light distribution, thus
the Balmer decrement extinction may not exactly match the extinction
of the stellar light.

If we assume that the emission is primarily powered by star-formation,
then we can estimate the metallicity based on strong-line indicators.
Using the metallicity calibration of the [OII]/[NII] ratio by
\citet{kewley02}, we find a metallicity of 12+log(O/H)=8.84, roughly
solar.  We can also estimate the gas-phase metallicity from the
\citet{tremonti04} metallicity-luminosity relationship; assuming $M_B
\sim -16.75$ \citep[using values from][]{tikhonov03}, we get a
metallicity 12+log(O/H)=8.33.  These results suggest a solar or
slightly sub-solar current gas-phase metallicity.

Finally, we can place an upper limit on the current star-formation
rate in the nucleus using the H$\alpha$ luminosity.  Correcting for
extinction, we get an H$\alpha$ flux of 10.8$\times
10^{-14}$~ergs$\,$s$^{-1}\,$cm$^{-2}$ corresponding to $L_{H\alpha} =
1.2 \times 10^{38}$~ergs$\,$s$^{-1}$.  This gives an upper limit to
the current star formation rate of $1.0 \times
10^{-3}$~M$_\odot\,$yr$^{-1}$ using the \citet{kennicutt98a} relation.
The value is an upper limit as some of the H$\alpha$ flux may
originate from AGN activity.  The radio continuum detection of 3~mJy
at 1.4~cm by \citep{delrio04} implies a SFR of $\sim$2$\times
10^{-4}$~M$_\odot\,$yr$^{-1}$ if it is the result of recent star formation.

The UV spectra of \citet{maoz98} in an $0\farcs86$ aperture also
suggests the presence of O stars in the nucleus.  We correct their
estimate of the number of O stars to a distance of 3~Mpc and for
extinction of $A_V = 0.5-1.0$.
This gives a range of 17-170 O stars, suggesting a total stellar mass
of 4000-40000~M$_\odot$ assuming a \citet{kroupa93} IMF.  The low end
of these estimates are consistent with the 10$^{3-4}$~M$_\odot$ of
recent stars implied by the H$\alpha$ and radio SFRs assuming
formation over the last $\sim$10~Myr.

\subsection{Nuclear Cluster Stellar Populations}

\begin{figure*}
\epsscale{1.2}
\plotone{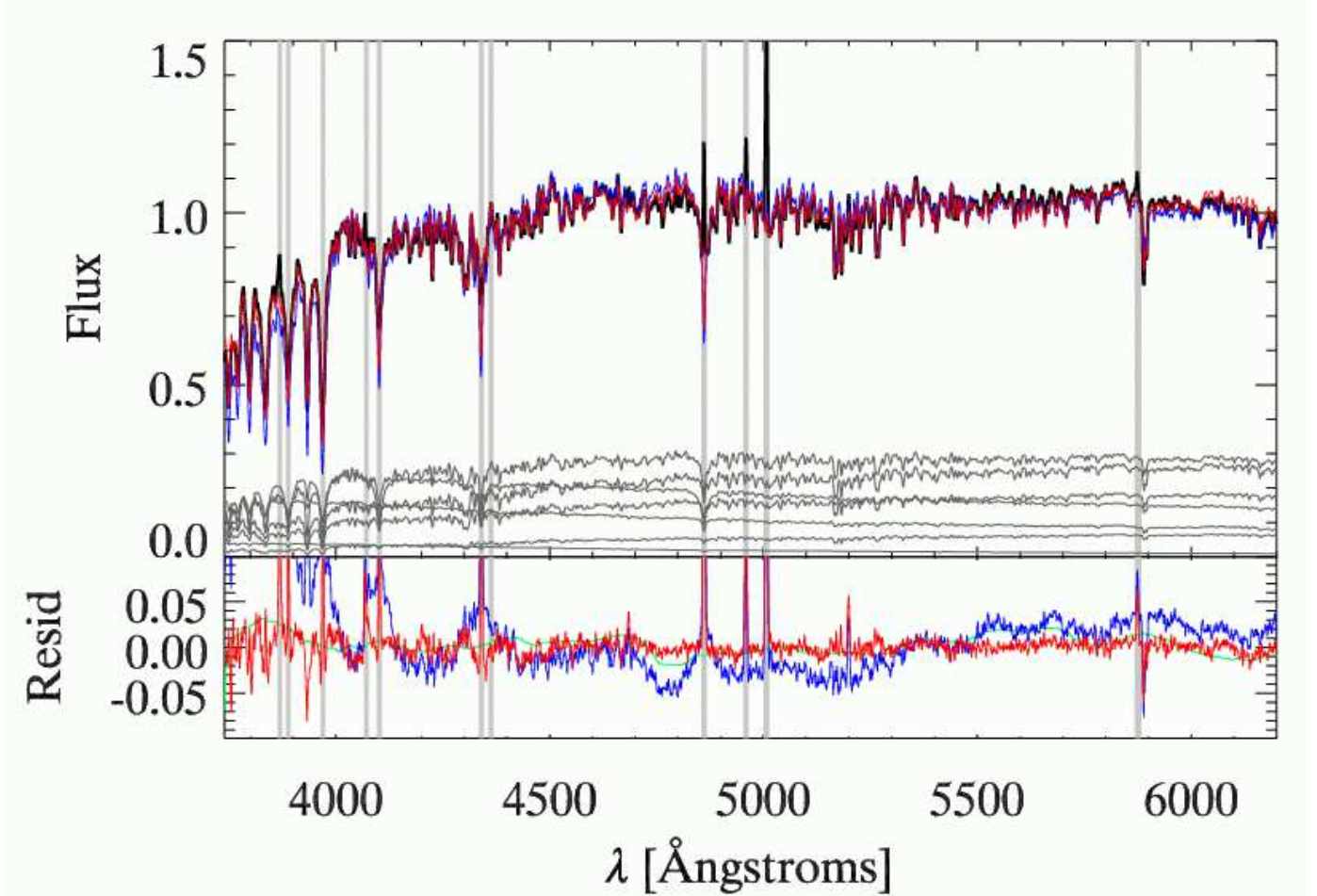}
\caption{{\it (Top panel --)} Normalized spectrum of the NGC~404
nucleus (black) with the best-fitting single age (blue) and multi-age
(red) fits to the spectrum. Vertical lines show emission line regions
that were excluded from the fit.  Gray spectra indicate the different
aged SSPs contributing to the multi-age fit.  {\it (Bottom panel) --}
residuals of the best fit single age (blue) and multi-age (red) fits.
Also shown is the flux calibration correction described in the text. }
\label{specfig}
\end{figure*}

% New Resub
In this section we fit the integrated nuclear optical spectrum to
determine the stellar populations of the NSC.  We find that stars with
a range of ages are present, but that stars 1-3~Gyr make up
$\gtrsim$50\% ($6\times10^5$~M$_\odot$) of the NSC mass.  This
dominant population has a solar or slightly subsolar metallicity
([Z]=-0.4).  About $10^4$~M$_\odot$ of young stars are also present.
These results are consistent with other studies of the NGC~404 nucleus
\citep{cidfernandes04,bouchard10}.

We fit the flux-calibrated spectrum between 3740-6200\AA, avoiding the
second order UV contamination at wavelengths above 6200\AA.  For
stellar population models we use both (1) the \citet{bruzual03} models
(BC03) which are based on Padova stellar models
\citep{bertelli94,girardi00} and the STELIB library of stellar spectra
\citep{leborgne03} and (2) the updated Charlot \& Bruzual 2007 (CB07)
version of these models incorporating the AGB models of
\citet{marigo07} and \citet{marigo08} obtained from St\'ephane
Charlot.  These models provide spectra for single age simple stellar
populations (SSPs) over a finely spaced age grid and at metallicities
of Z=0.0001, 0.0004, 0.004, 0.008, 0.02, and 0.05 at a spectral
resolution similar to our observations.  
%Added Resub
We find that typically the
CB07 models more accurately describe the data; they differ from the
BC03 models by up to 10\% across the wavelength range we are
considering here.

To remove known problems in the wavelength calibration of the STELIB
library \citep{koleva08} and any wavelength issues in our own
spectrum, we determined the relative velocity between our spectrum and
the SSP models every 100\AA.  Variations of up
to $\pm$35 km$\,$s$^{-1}$ around the systemic velocity were found and
corrected.  We also made a flux calibration correction of $\leq 3$\%
across the spectrum that was determined from residuals of our fits to
the nuclear clusters in NGC~404, NGC~205, M33, and NGC~2403 -- this is
described in more detail below.  We fit the models to our spectrum
using Christy Tremonti's {\em simplefit} program \citep{tremonti04},
which uses a Levenberg-Marquardt algorithm to perform a $\chi^2$
minimization to find the best-fit extinction \citep[using the
prescription of][]{charlot00} and scalings of the input model spectra.
The code excludes the regions around expected emission lines from the
fit.

We started our fitting by considering just single age stellar
populations.  Over the full range of 3740-6200\AA, the best-fitting
SSP is the CB07 model with Age=1.14~Gyr, Z=0.004, and an
$A_V=0.85$~mags.  The reduced $\chi^2$ is 21.8 corresponding to
typical residuals of 3\%.  This fit is shown in Fig.~\ref{specfig} as
the blue line.  The residuals are dominated by a mismatch between the
large-scale shape of the nuclear cluster continuum with the SSP model.
To mitigate these effects and get a better sense of the metallicity,
we also tried fitting just the metal line rich region from
5050-5700\AA.  The best fit is again a CB07 model with an age of 1.14
Gyr, but with Z=0.008 and $A_V=0.91$.  This metallicity and reddening
are quite similar to the values expected based on the emission lines
and the mass-metallicity relationship.

Previous studies have found significant evidence for multiple stellar
populations in nuclear star clusters
\citep{long02,rossa06,walcher06,seth06,siegel07}.  While most of these
studies have focused on spiral galaxies, there is also direct evidence
that the NGC~404 nuclear star cluster contains multiple stellar
populations; \citet{maoz98} finds evidence for massive O stars,
suggesting a population with age $\lesssim 10$~Myr, while based on the
color of the nucleus and its spectrum, older stars are clearly present
as well.  
%Added Resub
Also, as discussed in \S3.1, we
expect contamination of $\sim$30\% from the bulge in our MMT spectrum.

\begin{deluxetable*}{lccccccccccc}
\tablewidth{0pt} 
\tablecaption{Best-Fit NSC Stellar Population Model \label{nscpoptab}}
\startdata
\tableline
Age [Myr] & 1 & 10 & 50 & 101 & 286 & 570 &1015 &2500 &5000 & 10000 & 13000\\
Z & 0.02 & 0.02 & 0.02 & 0.02 &0.02 &0.02 &0.02 &0.02 & 0.008 & 0.008 & 0.0004\\
\tableline
Light Fraction & 0.013 & 0.090 & 0.000 & 0.000 & 0.157 & 0.000 & 0.277 & 0.239 & 0.166 & 0.058 & 0.000 \\
Mass Fraction & 0.000 & 0.003 & 0.000 & 0.000 & 0.035 & 0.000 & 0.152 & 0.317 & 0.316 & 0.177 & 0.000

\enddata

\end{deluxetable*}

We tried a number of approaches to our multi-age fits.  All of these
involved fitting a small subsample ($\sim$10) of the SSPs with ages
ranging from 1~Myr to 20~Gyr \citep[see][for a more complete
discussion of stellar population
synthesis]{walcher06,koleva08}.  We tried three types of models with both the CB07 and BC03 spectra:
\begin{enumerate}
\item Single-metallicity models using the 10 default ages in the {\it
simplefit} code (5, 25, 101, 286, 640, 904, 1434, 2500, 5000, 10000
Myr) or the 14 age bins used by \citet{walcher06} (1, 3, 6, 10,
30, 57, 101, 286, 570, 1015, 3000, 6000, 10000, 20000 Myrs).
\item Inspired by observations of the Sgr dSph nucleus \citep[a.k.a.
M54;][]{monaco05,siegel07} we also tried fitting models with chemical
evolution from low metallicity (Z=0.0004 / [M/H]=-1.7) at old ages
(13~Gyr) to solar metallicity for young ages \cite[following Fig.~2
from][]{siegel07}.  Specifically we used a model with Z=0.02 at ages
of 1, 10, 50, 101, 286, 570, 1015, and 2500 Myr plus Z=0.008 models at
5000 and 10000 Myr and Z=0.0004 models at 13000 Myr.  
\item Assuming a continuous or exponential star formation history with
chemical evolution similar to the previous item.  
\end{enumerate}

Of all the single metallicity models (model 1), the best fit is
obtained with the CB07 spectra using the default ages with Z=0.008
yielding a reduced $\chi^2$ of 4.3 and typical residuals of 1\%. This
is a vast improvement over the best fit SSP model.  The fit is
dominated by intermediate age stars, with 72\% of the light in the
1.434~Gyr SSP.  Many other models also produce nearly as good fits as
this one including models with solar and super-solar metallicity.

All the best-fitting ($\chi^2_{red} < 5$) models have fairly similar
age distributions with 50-85\% of the light coming from ages between
600 and 3000~Myrs, 5-6\% of the light coming from ages $<$10~Myr, and
0-25\% of the light coming from ages $>$5~Gyr.  The $A_V$ values range
from 0.84 to 1.16 and the implied $V-I$ colors range from 0.84-1.04.

The spectrum is not well fit by a constant star formation history
(model 3), giving a reduced $\chi^2$ of 22.  The constant SFH clearly
has too many young stars, with the continuum mismatched in a way that
cannot be compensated for by the reddening.  We note that a constant
SFH does provide good fits to the spectra of nuclei in many late-type
galaxies \citep{seth06,walcher06}.  An exponentially declining SFH
with a timescale of $\sim$8~Gyr does a better job of matching the
overall continuum of the observations, but still yields a reduced
$\chi^2$ double that of our best fitting models.

The best-fit model is obtained from model 2, which contains
metallicity evolution.  The reduced $\chi^2$ of the best fit is 4.23,
slightly better than any of the single metallicity models. This model
is shown in red in Fig.~\ref{specfig} with the individual SSP
components shown in gray in the top panel.  The best-fit model has
significant contributions from many of the individual ages, these are
shown in Table~\ref{nscpoptab}.  It is once again dominated by 1-3~Gyr
old stars, which make up half the mass of the NSC.  This appears to be
a robust result, with all our best-fit models containing a similar
result.  The luminosity weighted mean age is 2.3~Gyr, the mass
weighted mean age is 4.3~Gyr.  The best fit $A_V$ is 0.88 and the
$M/L$s are 0.89, 0.73, and 0.34 in the $V$-, $I$-, and
$H$-bands.  NGC~404 does not seem to have a dominant old metal-poor
population like that seen in M54 \citep{monaco05}.  From the $M/L$
analysis presented in \S5.3, we find that the maximum contribution for
an old (13 Gyr) metal-poor (Z=0.0004) population that keeps the
reduced $\chi^2$ below 5 is $\sim$17\% of the luminosity in the
$V$-band and $\sim$35\% of the total mass.

We can compare our best-fit model to other observed properties of the
nucleus.  The integrated colors of the best-fit model are $I-H$=1.59
and $V-I$=0.96.  An estimate of the $I-H \sim 1.7$
(Table~\ref{sersictab}) comes from the SB profiles, while the
$F547W-F814W \sim V-I \sim 0.95$ comes from the unreddened W portion
of the nucleus in Fig.~\ref{colorfig}.  Both of these quantities need
to be corrected for foreground extinction of $E(I-H)$ and $E(V-I)$ of
0.06 and 0.08 magnitudes; after this correction the measured
integrated colors agree with the best-fit spectral synthesis model to
within 10\%.

The best-fit model also suggests that the young stars ($\leq 10$~Myr)
make up 6\%/0.2\% of the light/mass in $H$-band and 8\%/0.3\% of the
light/mass $I$-band.  This gives a mass of of about $1.1 \times
10^4$~M$_\odot$ in young stars, in very good agreement with H$\alpha$
SFR integrated over 10~Myr and the estimates from the UV spectra
\citep[see discussion in \S5.1;][]{maoz98}.  

Finally, we compared the spectrum of NGC~404 to other nuclear star
clusters for which we took spectra on the same night, including M33,
NGC~205, NGC~2403, and NGC~2976.  Both M33 and NGC~205 do better than
any single stellar population model at matching the NGC~404 spectrum,
with reduced $\chi^2$ values of 9.3 and 15.0.  The stellar populations
in the M33 nucleus have been previously discussed by \citet{long02}.
They find the spectrum requires a large contribution of stars with an
age of $\sim$1~Gyr and a lesser contribution of stars at 40~Myr.  This
dominant 1~Gyr population is consistent with our findings above.  In
our initial fits of these nuclear star clusters we noted that the
residual pattern was quite similar regardless of the details of the
fit, with the most dramatic feature being a change of a few percent
between 4700-4800\AA.  Our flux calibration errors are expected to be
at this level based on the residuals from the flux calibrators.  Given
the similarities of the residuals in all of the nuclear spectra we
medianed the best fit model 1 residual spectra from NGC~205, M33,
NGC~404, NGC~2403, then smoothed it and used it as a correction to our
flux calibration.  The correction is shown as the green line in the
residual plot in Fig.~\ref{specfig}.  This change improved the
$\chi^2$ of the single-stellar population and multi-age fits
substantially at all metallicities but had little effect on the
best-fit parameters.

The stellar population of NGC~404 has been previously studied by
\citet{cidfernandes04} who analyzed the blue spectrum (3400-5500\AA)
of the NGC~404 nucleus as part of a study of a large number of LINER
nuclei.  Using empirical templates of five nuclei ranging from young
to old they find a best fit for NGC~404 of 78.5\% of light in an
intermediate age template (for which they used a spectrum of NGC~205)
and 21.5\% from old templates, with an $A_V$ of 0.90 and residuals of
6.6\%; this residual is much larger than for our fits (with $\sim$1\%
residuals) due primarily to the lower S/N of their spectrum.  Their
fit agrees quite well with our finding of a $\sim$1~Gyr old population
in the nucleus and our best-fit reddening.  Unlike us, they find no
evidence for young stars, probably because the NGC~205 template
spectrum used for intermediate ages also includes a young stellar
component \citep[e.g.][]{monaco09}.  The nuclear stellar populations
have also been recently studied by \citet{bouchard10} using a
different fitting code and stellar library.  Their best-fit
four-component model matches ours very well, with 4\% of the light in
stars younger than 150~Myr, 20\% at 430~Myr, 62\% at 1.7~Gyr, and 14\%
at 12~Gyr.

\subsection{Nuclear Cluster Mass to Light Ratio}

The multi-age population fits to the NGC~404 nuclear spectrum are
degenerate, with many possible combinations of ages and extinctions
giving similar quality fits.  This translates into an uncertainty in
the stellar population $M/L$, with the presence of more young stars
decreasing the $M/L$ and the presence of more old stars increasing it.
For comparison with the dynamical models (\S6.1), we'd like to
quantify the range of possible $M/L$s allowed by spectral synthesis
fits.

To try to quantify the possible variation in the $M/L$, we took
the best-fit model described in the previous section (model 2, with
CB07 and current solar metallicity) and fixed the age of each
individual SSP (e.g. just the 10 Gyr population) to have between
0-100\% of the flux of the unconstrained best-fit model, while the
extinction and contribution of the rest of the SSPs were allowed to
vary.  We then took all 'acceptable' fits (defined below) and examined
the range in $M/L$.  Because we have SB profiles in both $I$-
and $H$-bands (Fig.~\ref{sbfig}), we focus on the $M/L$s in
those bands.

We choose to consider all fits with reduced $\chi^2$ values of $< 5$
as acceptable, a value 18\% higher than our best fitting models.  This
choice is informed by the $\chi^2$ probability distribution; our fits
have $\sim$660 degrees of freedom and ignoring any model errors, the
99\% confidence limit for this distribtion is a $\Delta\chi^2$ of
13\%.  Unfortunately, since our best fit has a reduced $\chi^2$ is
4.2, there must be significant deficiencies in the models or flux
calibration at the 1\% level.  Nonetheless, the range of $M/L$s with
reduced $\chi^2 < 5$ should reflect the a conservative estimate of the
acceptable values, roughly equivalent to 3$-\sigma$ error bars.

We find that the range of acceptable $M/L$s is 0.68-1.22 in $I$ band
and 0.29-0.61 in $H$ band.  We note that the best-fit $M/L$ of 0.73 in
$I$ band and 0.34 in $H$ band is near to the lower range of these
values, suggesting a larger degeneracy in the mass of old stars than
for the younger populations.  Our best-fit values are consistent with
with the dynamical $M/L_I=0.70\pm0.04$ derived in \S6.1.

\subsection{Bulge Stellar Populations}

In \citet{cidfernandes05}, they analyze the radial variations in the
NGC~404 spectrum and find increasing age with radius, with the old and
intermediate age templates having equal weight at radii between 5 and
10 arcseconds.  We use our MMT spectra of NGC~404 to extract a spectra
at radii between 5\asec~and 10~\asec from the nucleus.  From our
profile fits (Fig.~\ref{sbfig}), we expect that the bulge component
should dominate this spectrum.  This spectrum has a peak S/N of 70
around 5000\AA~dropping to 25 at 3700\AA.

The models that best fit the NGC~404 bulge spectrum are the same sets
of models as the nuclear spectrum (i.e.~model 1 with Z=0.008 
and model 2).  However, like \citet{cidfernandes05} we find that the
bulge has a significantly older population with all the best fits
having $>$50\% of the light in an old ($\geq 5$~Gyr) component.  The
best fits have $A_V$ between 0.16 and 0.24 which corresponds well to
the $A_V$ of 0.19 expected for the foreground extinction
\citep{schlegel98} and is consistent with the dust absorption being
primarily restricted to the inner 5\asec of the galaxy as suggested by
HST color images.  The best fits all have residuals of $\sim$2.6\%,
corresponding to reduced $\chi^2$ of 2.0.  All the fits have $\sim$2\%
of the light in a young ($<$100~Myr) component and a significant
contribtution ($\sim$30\%) from a $\sim$1~Gyr component.  This
analysis clearly shows a significant population difference between the
bulge and nuclear spectrum which supports the separation of the SB
profiles into separate components in \S3.

As all the models have nearly identical fits, we infer $M/L$ ratio
using model 2, which best fit the nuclear star cluster spectrum.  We
find a $M/L$ of 1.69, 1.28, and 0.63 in the $V$-, $I$-, and $H$-bands.

Two contemperaneous papers suggest that a dominant old population
extends out to larger radii as well.  At $r \sim 25$\asec,
\citet{bouchard10} find two-thirds of the mass is an old component
using spectral synthesis.  And using a deep color-magnitude diagram at
a radius of $\sim$160\asec (near the disk-bulge transition),
\citep{williams10} have found $\sim$80\% of the stellar mass is
$>$10~Gyr old.  The nucleus, with its dominant intermediate age
population, is therefore quite distinct from the rest of the galaxy.

\section{Dynamical Modeling}

In this section we present dynamical modeling of the central mass
distribution of NGC~404 using both the stellar (absorption line) and
gas (H$_2$ emission) kinematics.  From modeling of the stellar
kinematics, we find a $M/L_I = 0.70 \pm 0.04$.  We can use this to
derive a total mass for the nuclear star cluster by combining this
estimate with the total luminosity of the NSC and central excess
(Table~2) to get a mass of $1.1 \pm 0.2 \times 10^7$~M$_\odot$.  The
error on the NSC mass includes additional uncertainty due to the
possible variation in $M/L$ of the central excess.  We note that the
dynamical $M/L_I$ matches the best-fit $M/L_I$ derived from the
stellar population fits to within 10\%.

We can also constrain the presence of a MBH using the dynamical
models.  The two methods give inconsistent results for the BH mass,
with the stellar kinematics suggesting an upper limit of $<1 \times
10^5$~M$_\odot$, while the gas models are best fit with a MBH of mass
$4.5^{+3.5}_{-2} \times 10^5$~M$_\odot$ (3$\sigma$ errors).  These
measurements assume a constant $M/L$ within the NSC; taking into
account possible variations of $M/L$ at very small radii, the BH mass
estimates/upper limits can increase by up to $3 \times
10^5$~M$_\odot$.  If there is a MBH in NGC~404, its mass is
constrained to be $< 10^6$~M$_\odot$, smaller than any previous
dynamical BH mass measurement in a present-day galaxy.

\subsection{Stellar Dynamical Model}

\begin{figure}
  \plotone{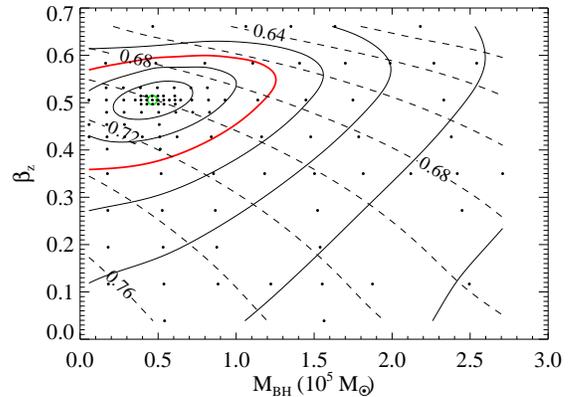}
  \caption{$\chi^2$ contours describing the agreement between the JAM
  dynamical models and the NIFS integral-field observations of $V_{\rm
  rms}$. The models optimize the three parameters $M_{\rm BH}$,
  $\beta_z$ and $M/L$ but the contours are marginalized over the
  $M/L$, which is overplotted as dashed contours with labels.  Dots
  show the grid of models. The three lowest $\chi^2$ contour levels
  represent the $\Delta\chi^2=1$, 4 and 9 (thick red line)
  corresponding to 1, 2, and $3\sigma$ confidence levels for one
  parameter.}
  \label{fig:chi2}
\end{figure}

\begin{figure}
  \plotone{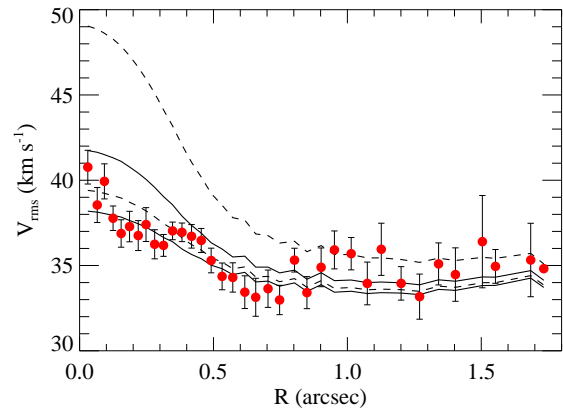}
  \caption{The observed NIFS profiles of $V_{\rm rms}$, biweight
  averaged over circular annuli (data points), are compared to JAM
  model predictions of the same quantity, for different $M_{\rm BH}=0$
  (lowest solid line), 1, 3, 5, and 10$\times10^5$~M$_\odot$ (top
  dashed line). All models have $M/L_I=0.70$ $M_\odot/L_\odot$ and
  $\beta_z=0.5$ as the best fitting solution.}
  \label{fig:jam}
\end{figure}

We constructed a dynamical model to estimate the mass and $M/L$ of the
nuclear star cluster and of a possible central supermassive black hole
inside it.  The first step in this process is developing a model for
the light distribution.  To parametrize the surface brightness
distribution of NGC~404 and deproject the surface brightness into
three dimensions, we adopted a Multi-Gaussian Expansion
\citep[MGE;][]{emsellem94}. The MGE fit was performed with the method
and software\footnote{Available from
http://purl.org/cappellari/idl/\label{jam}} of \citet{cappellari02}.
We fit a 2-D form of the $I$-band data as presented in \S3: at $r <
5$\asec, we use the WFPC2 $I$-band image with reddened regions masked
out, while outside of 5\asec, we use the ACS data (which is saturated
in the nucleus).  Although the use of NIR data would be preferable
given the presence of dust in the nucleus, the uncertainty in the NIFS
PSF and the poor resolution and sampling of the available HST NICMOS
data makes the F814W data the best available image.  The MGE fit takes
into account the PSF: we use a Tiny Tim PSF \citep{krist95}.  In the
fit we forced the observed axial ratios $q'$ of the MGE Gaussians to
be as big as possible, not to artificially constrain the allowed
inclination of the model. The PSF-convolved MGE represents a
reasonable match to the $F814W$-band images with some mismatch seen in
the inner parts due to dust.  We note that modeling results based on
NICMOS F160W mass models are consistent with those presented here.

The primary uncertainty in the determination of $M_{\rm BH}$ results
from uncertainties in the central stellar mass profile.  The $F814W$
luminosity at $r < 0\farcs05$ (0.7~pc) is $\sim 1.3 \times
10^6$~L$_\odot$, while the MGE model has a luminosity of $\sim 4 \times
10^5$~L$_\odot$ within a sphere of the same radius.  Translating this
luminosity into a stellar mass density is complicated given the
possible presence of dust, stellar population gradients, non-thermal
emission and possible nuclear variability.  Furthermore, the available
kinematics sample only the very nuclear $R\la1\farcs5$ region of the
galaxy. This spatial coverage is not sufficient to uniquely constrain
the orbital distribution and $M_{\rm BH}$ in NGC~404
\citep{shapiro06,krajnovic09}.  This complex situation does not
justify the use of general brute-force orbital superposition dynamical
modeling methods \citep[e.g.][]{richstone88,vandermarel98}. This
implies we will not be able to uniquely prove the existence of a BH in
this galaxy, but we can still explore the ranges of allowed $M_{\rm
BH}$ by making some simplifying observationally-motivated
assumptions. A similar, but less general approach was used in other
studies of BHs in NSCs from stellar kinematics
\citep{filippenko03,barth09}.  We note that because we kinematically
resolve the NSC and can determine its $M/L$, we can provide stricter
BH upper limits than those provided by these papers, in which the NSC
kinematics were unresolved.

The adopted dynamical model uses the Jeans Anisotropic MGE (JAM)
software\footnotemark[10] which implements an axisymmetric
solution of the Jeans equations which allow for orbital anisotropy
\citep{cappellari08}. Once a cylindrical orientation for the velocity
ellipsoid has been assumed, parameterized by the anisotropy parameter
$\beta_z=1-\sigma_z^2/\sigma_R^2$, the model gives a unique prediction
for the observed second velocity moments $V_{\rm
rms}=\sqrt{V^2+\sigma^2}$, where $V$ is the mean stellar velocity and
$\sigma$ is the velocity dispersion. The JAM models appear to provide
good descriptions of integral-field observations of large samples of
real lenticular galaxies \citep{cappellari08,scott09} and give
similar estimates for BH masses as Schwarzschild models
\citep{cappellari10}.

The luminous matter likely dominates in the extreme high-density
nucleus of NGC~404. For this dynamical model we assume light traces
mass with a constant $M/L$. For the model we assumed a nearly face-on
inclination of $i=20^\circ$ ($i=90^\circ$ being edge on) as indicated
by the regular HI disk kinematics and geometry \citep{delrio04}, and
further supported by the nearly circular isophotes at small
radii. However, unlike the gas dynamical model, our results do not
depend strongly on this inclination assumption.  The model has three
free parameters: (i) The anisotropy $\beta_z$, (ii) the mass of a
central supermassive black hole $M_{\rm BH}$ and (iii) the $I$-band
total dynamical $M/L$. To find the best fitting model parameters we
constructed a grid for the two nonlinear parameters $(\beta_z,M_{\rm
BH})$ values and for each pair we linearly scaled the $M/L$ to match
the data in a $\chi^2$ sense. The resulting contours of $\chi^2$,
marginalized over $M/L$ are presented in Fig.~\ref{fig:chi2}. They
show that with the assumed stellar mass distribution, the models
suggest a $M_{\rm BH} < 1 \times 10^5$~M$_\odot$ with $\beta_z \sim
0.5$ and $M/L=0.70\pm0.04$ $M_\odot/L_\odot$ in $I$-band ($3\sigma$
levels or $\Delta\chi^2=9$).  A corresponding data-model comparison is
shown in Fig.~\ref{fig:jam}.  The dynamical $M/L_I$ (which includes a
correction for foreground extinction) is quite close to our best-fit
stellar population estimate of $M/L_I = 0.73$.

\begin{figure*}
\plotone{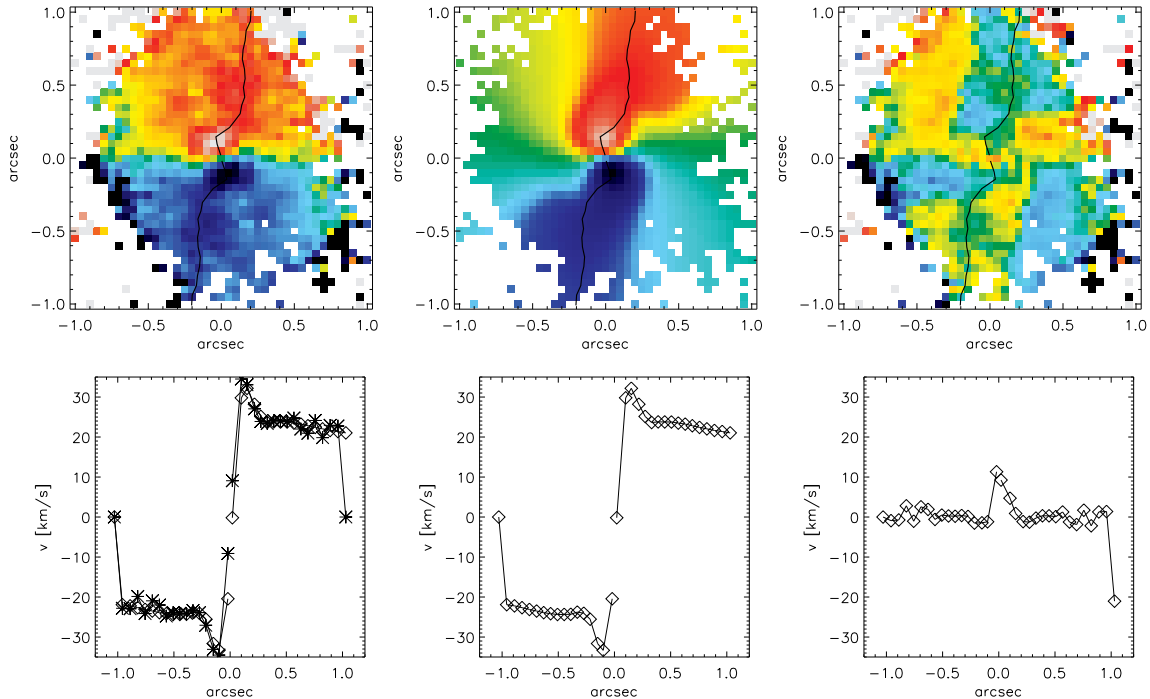}
\caption{Velocity field of the best-fitting H$_2$ dynamical model
(middle, top), with a black hole mass of $M_{\rm BH} = 4.5 \times
10^5$M$_{\odot}$ and a disk inclination of $37^\circ$, in comparison
to the symmetrized data (top left). The velocity residual (data-model)
is shown in the right panel. The velocity curves in the bottom panels
are extracted along the line of nodes (overplotted to the velocity
maps), and represent the peak velocity curves. The diamonds correspond
to the model velocity curve, while the crosses correspond to the
data.}
\label{fig:gas_model}
\end{figure*}

As noted above, the largest uncertainty in our MBH measurements comes
from uncertainty in the central stellar mass density hidden by the
assumption of a constant $M/L$.  The bluer color in the $F547M-F814W$
color image (Fig.~\ref{colorfig}) and the presence of UV emission at
$r \lesssim 0\farcs05$ suggests a possible decrease in $M/L$ which
could mask the presence of a possible BH.  Because our optical
spectroscopy does not resolve the central stuctures, we can only use
the HST imaging data to try constraining possible variations in the
$M/L$.  We now determine a limit on how much the central $M/L$ may
vary by quantitatively examining the color difference between the NSC
and the central pixels.  We measured the NSC colors just to the W of
the nucleus and find a $F330W-F547M$ ($U-V$) of 0.48 and an
$F547-F814W$ ($V-I$) of 0.87 after correction for foreground
reddening.  The colors at $r < 0\farcs05$ are both about 0.1
magnitudes bluer than the NSC colors.  In $V-I$ the presence of a
$\sim$10~Myr population will give the largest reduction in $M/L$ with
the least affect on the color, since these populations are dominated
by red supergiants at long wavelengths \citep[e.g. see Fig.~4
in][]{seth08b}.  For a solar metallicity population, the $V-I$ color
at 10~Myr is about 0.5 \citep{girardi00}, meaning the observed nuclear
color difference is consistent with about 30\% of the light in young
stars.  This would result in a $\sim$30\% reduction in the $M/L$, as
the $M/L$ for a 10~Myr population is very small.  However, this large
fraction of a young population would imply a larger change in $U-V$
color than observed, so 30\% represents an upper limit on the
reduction in $M/L$.  If we take the $I$-band luminosity at $r <
0\farcs05$ of $1.3 \times 10^6$~L$_\odot$ with our default $M/L$ of
0.70, we get a mass at $r < 0\farcs05$ of $9 \times 10^5$~M$_\odot$.
Reducing the $M/L$ by 30\% at the center relative to the NSC would
therefore reduce the central stellar mass {\it and increase the BH
mass} by $\sim 3 \times 10^5$~M$_\odot$.  Therefore, even including
the effects of a possible decline in the central $M/L$, the BH mass is
constrained to be $\lesssim 4 \times 10^5$~M$_\odot$ by the stellar
dynamical model.  Future HST spectroscopy that can resolve the stellar
populations gradient in the nucleus should produce a more accurate
mass model that will enable us to robustly test the presence of a MBH
in NGC~404.

\subsection{Gas Dynamical Model}

\begin{figure}
\plotone{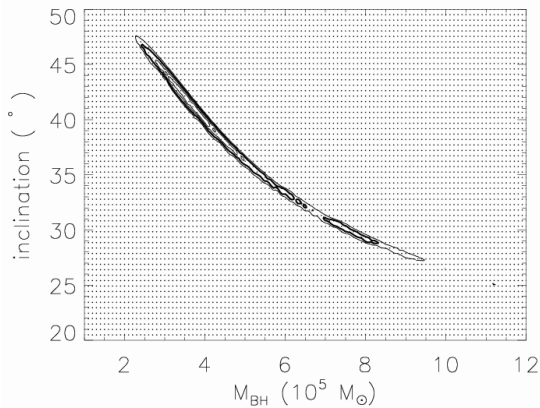}
  \caption{Constraining the mass of the central BH: the figure
indicates the grid of models (in black hole mass, $M_{\rm BH}$, and
disk inclination) that was calculated, and the contours show $\chi^2$
in the vicinity of the best-fit dynamical models for matching the
H$_2$ kinematics. The minimum $\chi^2$ model is at $M_{\rm BH}\sim 4.5
\times 10^5$M$_{\odot}$ and a disk inclination of $37^\circ$. The
contours indicate the 1, 2, and 3 $\sigma$ confidence levels,
respectively (see text for details).}
  \label{fig:gas_chi2}
\end{figure}

In addition to the stellar dynamical model we constructed a dynamical
model to reproduce the kinematics of molecular hydrogen emission
(\S4.2) to obtain an additional constraint on the presence of a
possible central BH.  We follow the modeling approach of
\citet{neumayer07} that assumes the gas to be settled in a thin disk
in the joint potential of the stars and a putative BH. We assume the
gas is moving on circular orbits and we do not include a pressure term
in the gas model, as the velocity dispersion of the gas is small. We
have tested the effect of an additional pressure term and find it to
be very small, well below our derived errors.  The stellar potential
is fixed using the MGE parameters and the stellar $M/L_I$ derived
\S5.2, and thus suffers from the same uncertainty in the central
stellar mass profile.

The gas velocity map appears quite complex with sudden changes in
position angle of the major kinematic axis and some of the kinematic
features might be due to non-gravitational motions driven by the LINER
nucleus.  However, as described in \S4.2, there also appears to be
clear rotation in the central $0\farcs4\times0\farcs4$ with a major
rotation axis of $\sim 15^\circ$, and a twist to slightly negative PAs
at larger radii. We interpret this twist in the line-of-nodes as a
change in inclination angle of the molecular gas disk and model the
kinematics using a tilted ring model. The inclination angle of the gas
disk for radii $>0\farcs2$ is fixed to the value of $20^\circ$ that is
also used for the stellar model and consistent with inclinations
derived by \citet{delrio04} from HI. The dynamical model has two free
parameters: (i) The inclination angle of the inner (r$\le 0\farcs2$)
H$_2$ disk $i$ and (ii) the mass of a central supermassive black hole
$M_{\rm BH}$. To find the best fitting model parameters we constructed
a grid for the two parameters ($i$,$M_{\rm BH}$) and matched the data
(velocity and velocity dispersion) minimising
$\chi^2$. Figure~\ref{fig:gas_chi2} shows the contours of $\chi^2$ and
indicates the best fit model parameters $i=37^\circ \pm10^\circ$ and
$M_{\rm BH}= 4.5^{+3.5}_{-2.0} \times 10^5$ M$_\odot$ (3-$\sigma$
errors). The best-fitting model for the H$_2$ velocity field is
compared to the data in Fig.~\ref{fig:gas_model}.  We note that the
best-fit inclination is consistent with the axial ratios of the
emission (\S4.2).

As Figure~\ref{fig:gas_chi2} shows, the inclination angle $i$ and the
central mass $M_{\rm BH}$ are strongly coupled, since the amplitude of
the rotation curve is proportional to $\sqrt{\mathrm{M_{BH}}} \times
\sin(i)$. This degeneracy in the gas model leads to the relatively
large uncertainties in inclination angle and black hole mass. However,
the gas kinematics cannot be reproduced without a central black hole,
even for an almost edge-on central inclination angle.  The
best-fitting black hole mass is significantly different from the
non-detection of $<1 \times 10^5$~M$_\odot$ derived from the stellar
kinematics.  One possible source of this discrepancy would be
non-gravitational motions that are not accounted for by the
gas-kinematical model.  Another possible problem is the assumption of
a thin gas disk, although thicker gas distributions provided worse
fits to the gas kinematic data.  We note that both determinations use
the same MGE mass model, and thus the uncertainty due to the central
$M/L$ applies to these results as well.  This could increase the BH
mass estimates by as much as $3 \times 10^5$~M$_\odot$.

%\newpage

\section{Discussion}

\subsection{Nuclear Star Cluster: Scaling Relationships and Formation}

\begin{figure}
\plotone{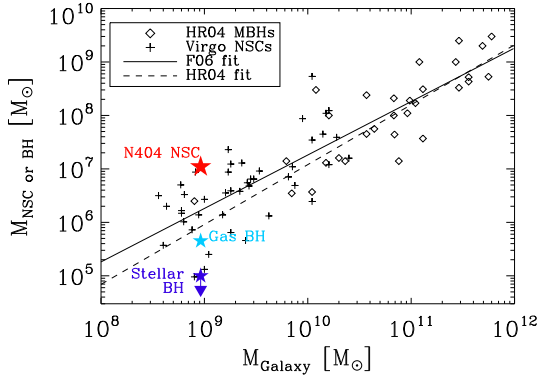}
\caption{Scaling relations for NSCs and MBHs in early-type galaxies.
Data for MBHs is from \citet{haring04}, while for the NSCs it is from
\citet{cote06} with mass estimates from \citet{seth08}. The solid line
shows the joint MBH/NSC fit from \citet{ferrarese06}, while the dashed
line shows the fit to the MBH data from \citet{haring04}. Just the
bulge mass is plotted for NGC~404, the total galaxy mass would be
$\sim$30\% larger.}
\label{scalingfig}
\end{figure}

We compare the NGC~404 NSC to the scaling relationships between
galaxy/bulge mass and BH/NSC mass in Fig.~\ref{scalingfig}.  We
have derived the mass of the NSC using the dynamical $M/L$ determined
from the NIFS stellar kinematics and the luminosity of the NSC (and
central excess) above the underlying bulge.  The NGC~404 NSC mass is a
large fraction (1-2\%) of the bulge/galaxy mass, making it fall
significantly above the relationship for MBHs and early-type Virgo
NSCs given by \citet{ferrarese06}, however, there are several other
similar outliers in the Virgo NSC sample.  Using the
\citet{ferrarese06} NSC mass vs. $\sigma$ relation also gives a mass
nearly an order of magnitude smaller than the measured NSC mass.

The NSCs high mass relative to its galaxy raises the question of
whether our interpretation of this component as an NSC is correct.  As
an alternative, we could be seeing some form of ``extra-light
component'' \citep{hopkins09a,kormendy09} or an ``extended nuclear
source'' \citep{balcells07} that are typically seen on somewhat larger
scales than NSCs in coreless elliptical, lenticular, and early-type
spirals.  The mass fraction of $\sim$1\% places the NGC~404 NSC in the
regime of overlap between NSCs and the extra-light components
\citep[Fig.~46 of][]{hopkins09a}.

\begin{figure}
\plotone{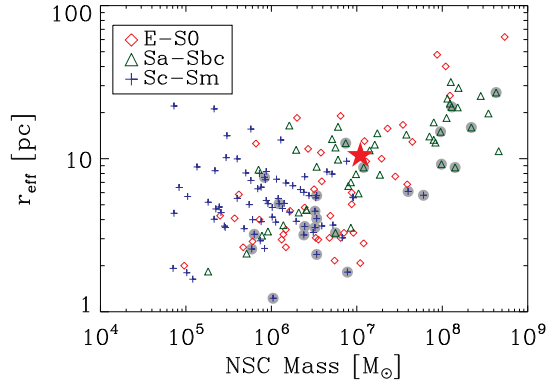}
\caption{Mass vs. size for NSCs collected in \citet{seth08}.  The
large solid star shows the position of the NGC~404 NSC, showing that it
has a size and mass typical of other early type galaxies.  Data is
from \citet{carollo97,carollo98,carollo02}, \citet{boker02},
\citet{cote06} and \citet{seth06}.  Points with underlying gray dots
have dynamical or spectroscopic $M/L$ estimates from \citep{walcher05}
and \citet{rossa06}.}
\label{sizemassfig}
\end{figure}

Using data collected in \citet{seth08}, we plot the size and mass of
clusters in Fig.~\ref{sizemassfig}, and find that the NSC in NGC~404
is typical of those in other early-type galaxies.  The NGC~404 NSC is
also typical in the mass-density plane, and the NSC S\'ersic index of
$\sim$2 is similar to that found for the M32 and Milky Way NSC in
\citep{graham09}.  The morphology of the NGC~404 NSC is therefore
completely consistent with other NSCs.  Its mass and density is also
similar to the lowest mass extra-light components
\citep[][Fig.~45]{hopkins09a}, and follows the size-surface brightness
scaling of the extended nuclear components \citep[Fig.~3
of][]{balcells07}.  In \citet{hopkins09a}, they argue that while the
extra-light and NSCs can overlap in morphological properties, the two
can still be distinguished based on their kinematics and stellar
populations.  We disagree with their argument.  Kinematically, both
NSCs and extra-light can be disky and rotating
\citep{krajnovic08,seth08b,trippe08} and in early-type galaxies, the
stellar populations of many NSCs are not significantly different from
their underlying galaxy \citep{cote06,rossa06}.  We therefore suggest
that the distinction between extra-light components and NSCs may be
ambiguous in some cases such as NGC~404.

The extra-light components in elliptical galaxies are thought to
result from gas funneled to the center of the galaxy during gas-rich
mergers \citep[e.g.][]{mihos94,hopkins09a}.  Given the complicated
kinematics and stellar population of the NGC~404 NSC, it must be
formed from multiple episodes of nuclear accretion.  However, it
formed about half its stars 1-3~Gyr ago.  In \citet{delrio04}, they
suggest NGC~404 underwent a merger with a gas-rich dwarf $\sim$1~Gyr
ago to explain the presence of the HI gas disk observed at large
radii.  The age estimate of the merger was based on the radius beyond
which the gas had not yet settled onto the galaxy plane.  It is
therefore plausible that the $\sim 6 \times 10^6$~M$_\odot$ of stars
formed at about this same age in the NSC resulted from accumulation of
merger gas at the center of the galaxy.  A merger origin is also
consistent with the presence of counter-rotation in the central
excess.  This scenario is quite different from the formation mechanism
suggested by observations of late-type galaxies
\citep{seth06,walcher06,seth08b}, which appear to form primarily by
episodic accretion of material (gas or stars) from the galaxy disk.
Resolved observations of a larger sample of NSCs in early-type
galaxies is required to determine if NGC~404 is an unusual case, or if
NSCs in early-type galaxies typically show some evidence gas accreted
from mergers.

\subsection{The Possible NGC~404 MBH}

Our dynamical modeling in \S6 provides mixed evidence for the presence
of an MBH in NGC~404 with mass less than 10$^6$~M$_\odot$ (making it
an IMBH candidate).  If verified through follow-up observations, this
BH would be the lowest mass dynamically detected BH at
the center of a galaxy, with previous determinations of low mass BHs
in M32 \citep[$2.5 \pm 0.5 \times 10^6$~M$_\odot$; ][]{verolme02} and
Circinus \citep[$1.7 \pm 0.3 \times 10^6$~M$_\odot$;][]{greenhill03}
being significantly more massive.  A reverberation mapping mass of the
BH in nearby Seyfert~1 galaxy NGC~4395 \citep{peterson05}, and
indirect mass measurements in other low mass Seyfert~1 galaxies
\citep[e.g.][]{greene07b} provide additional evidence for the presence
of IMBHs at the centers of lower mass galaxies.  Possible IMBHs at the
center of star clusters (and possible former nuclei) G1
\citep{gebhardt05}, $\omega$~Cen \citep{noyola08,vandermarel09}, and
M54 \citep{ibata09} suggest that IMBHs with masses of
$10^{4-5}$~M$_\odot$ may be widespread, but these results remain
controversial.  Measuring the occupation fraction and mass of BHs in
galaxies like NGC~404 is key to understanding the formation of MBHs in
general \citep[][]{volonteri08}.

From the latest version of the $M_{BH}-\sigma$ relations for different
galaxy samples from \citet{gultekin09}, the predicted $M_{BH}$ ranges
from $0.8-3 \times 10^5$~M$_\odot$ for a bulge $\sigma$ of
35-40~km$\,$s$^{-1}$.  This range of masses is consistent with all our
dynamical BH mass estimates.  A somewhat higher mass ($8 \times
10^5$~M$_\odot$) is implied by the \citet{haring04} relationship
(see Fig.~\ref{scalingfig}) and is consistent only with the gas-dynamical
estimate, while the \citet{ferrarese06} relationships have masses
$>10^6$~M$_\odot$ and are thus inconsistent with our data.  If a
black hole with a few $\times10^5$~M$_\odot$ resides in the center of
NGC~404, better knowledge of the mass model from future HST observations
should enable us to detect it.

The relative mass of NSCs and BHs within a single galaxy can probe the
relationship of these two objects.  \citet{seth08} find that the
relative mass of BHs and NSCs is of order unity (with a scatter of at
least one order of magnitude), while \citet{graham09} suggest a
possible evolution in the relative mass with the least massive
galaxies being NSC dominated, and the most massive galaxies being BH
dominated.  The BH in NGC~404 is $<$10\% of the NSC mass and thus is
not an obvious outlier from other galaxies where the mass of both
objects is known.

Finally, we note that the evidence of variability in the UV spectrum
\citep{maoz05} and possibly the NIR dust emission (\S3.1), suggests
that the NGC~404 black hole accretion is variable.  In both cases, the
observed flux decreased from observations taken in the 1990s to those
taken in the last decade.  This might help explain the ambiguous
nature of the AGN indicators in the NGC~404 nucleus; emission that
originates near the BH (e.g. hard X-ray emission, broad H$\alpha$) may
have recently disappeared, while the narrow-line region located at tens
of parsecs, may continue to show signs of the earlier accretion.
However, the observed hot dust emission suggests that the BH may still
be accreting at a very low level.

\section{Conclusions and Future Prospects}

This paper is the second resulting from our survey of nearby nuclear
star clusters, and demonstrates the rich detail we can obtain for
these objects.  The NGC~404 nucleus is a complicated environment, with
both a nuclear star cluster and a possible black hole.  We have found
that the surface brightness profile of the inner part of NGC~404
suggests the galaxy can be broken into three components: (1) a bulge
that dominates the light beyond 1\asec, with $M_{bulge} \sim 9 \times
10^8$~M$_\odot$, $r_{eff} = 640$~pc, and a S\'ersic index of
$\sim$2.5, (2) a NSC that dominates the light in the central arcsecond
with $r_{eff}=10$~pc and a dynamical mass of $1.1 \pm 0.2 \times
10^7$~M$_\odot$, and (3) a central excess at $r < 0\farcs2$, composed
of younger stars, dust emission, and perhaps AGN continuum.  NIFS IFU
spectroscopy shows that the NSC has modest rotation along roughly the
same axis as the HI gas at larger radii, while the central excess
counter-rotates relative to the NSC.  Furthermore, molecular gas
traced by H$_2$ emission shows rotation perpendicular to the stellar
rotation.  A stellar population analysis of optical spectra indicate
that half of the stars in the NSC formed $\sim$1~Gyr ago.  Some
ancient and very young ($<$10~Myr) stars are also present.  This star
formation history is dramatically different from the rest of NGC~404,
which is dominated by stellar populations $>$5~Gyr in age.  We suggest
a possible scenario where the burst of star formation in the NSC
$\sim$1~Gyr ago resulted from the accretion of gas into the galaxy
center during a merger.  This formation scenario is quite different
from the episodic disk accretion suggested by observations NSCs in
late-type galaxies.

Our dynamical modeling of the stellar and gas kinematics provide mixed
evidence for the presence of a black hole in NGC~404.  Assuming a
constant $M/L$ within the nucleus, the stellar dynamical model
suggests an upper limit of $1 \times 10^5$~M$_\odot$, as well as
measuring a $M/L_I = 0.70 \pm 0.04$ for the NSC.  The gas kinematics
are best fit by models including the presence of a black hole with
$M_{\rm BH}= 4.5^{+3.5}_{-2.0} \times 10^5$ M$_\odot$.  Both dynamical
BH mass estimates rely on a model for the stellar mass that we
construct from HST $F814W$-band imaging.  
%edited/added Resub
Uncertainties in the light profile (due to variability) and $M/L$ (due
to stellar population changes) within the central excess are of the
same order as the difference between the two black hole mass
estimates.  We have proposed to measure the mass model by using
HST/STIS spectroscopy to resolve the stellar populations within the
nucleus and additional multi-band imaging to extend this model to two
dimensions.  If successful we will combine this mass model with the
kinematic observations presented here as well as larger-scale
kinematics obtained from the MMT to more robustly determine the BH
mass.

In addition to the direct evidence of the BH from the dynamical
models, we find two other properties of the nucleus which suggest the
presence of an AGN.  First, we find unresolved hot dust emission at
the center of the NSC with a luminosity of $\sim1.4 \times
10^{38}$~ergs$\,$s$^{-1}$.  Comparison of the NIFS light profile to
previous NICMOS observations suggests that this dust emission may be
variable as is seen in other AGN.  Second, the H$_2$ line ratios
within the central arcsecond indicate thermal excitation in dense gas
similar to what is seen in other AGN.
%Added Resub
Our proposed HST observations request multi-epoch UV through NIR
imaging to search for definitive evidence of BH accretion in NGC~404.

Our nearby NSC survey includes 13 galaxies within 5~Mpc with $M_B$
between -15.9 and -18.8 for which we will have comparable data to that
presented here obtained using MMT, VLT, and Gemini.  These galaxies
span a wide-range of Hubble types in which we can examine the process
of NSC formation; there are four early type E/S0 galaxies and 9
spirals of type Sc and later.  NGC~404 represents one of the stronger
cases for finding a black hole, given its LINER emission and
proximity.  However, we expect to be able detect or place upper limits
of $\lesssim10^5$~M$_\odot$ on black holes in each of the sample
galaxies.

\vspace{0.1in}

\noindent {\em Acknowledgments:} We thank the referee, Jenny Greene,
for helpful comments, St\'ephane Charlot for sharing his models,
Christy Tremonti for sharing her code, the staff at Gemini and MMT,
NED, and ADS.  AS acknowledges support from the Harvard-Smithsonian
CfA as a CfA and OIR fellow, and helpful conversations with Margaret
Geller, Pat C\^ot\'e, and Davor Krajnovi\'c.  MC and NB acknowledge
support from STFC Advanced Fellowships. NN acknowledges support from
the Cluster of Excellence ``Origin and Evolution of the Universe".
Partially based on observations obtained at the Gemini Observatory,
which is operated by the Association of Universities for Research in
Astronomy, Inc..  Gemini data was taken as part of program
GN-2008B-Q-74.

{\it Facilities:} \facility{Gemini:Gillett (NIFS/ALTAIR)},
\facility{HST (ACS/WFC)}, \facility{MMT}

%\bibliography{/Users/seth/allreferences}
%\bibliographystyle{apj}

\end{document}